\documentclass[aps,pra,10pt,twocolumn,superscriptaddress,notitlepage,showpacs,floatfix,longbibliography]{revtex4-2}

\usepackage{bbm, amsthm, bm,textcomp, nicefrac,geometry,ragged2e}
\usepackage{enumitem}
\usepackage[dvipsnames]{xcolor}
\usepackage{float}
\usepackage[bbgreekl]{mathbbol}
\usepackage{graphicx,epstopdf,color,verbatim,enumitem}
\usepackage[normalem]{ulem}
\usepackage{pbox,array}
\usepackage{booktabs}
\usepackage{mathrsfs}
\usepackage{amssymb}
\usepackage{textgreek}
\usepackage{mathtools}
\usepackage{stackrel}
\usepackage[thinlines]{easytable}
\usepackage{amsmath}
\usepackage[utf8]{inputenc}
\makeatletter
\usepackage{soul}
\usepackage[caption=false]{subfig}
\usepackage{hyperref}
\hypersetup{
    colorlinks = true,
    linkcolor = teal,
    citecolor = teal,
    filecolor = Black,
    urlcolor = Black,
}
\usepackage[T1]{fontenc}
\usepackage[caption=false]{subfig}
\usepackage{babel}
\usepackage[linesnumbered,ruled]{algorithm2e}

\usepackage{braket}%Dirac Notation in QM
\usepackage{enumerate}
\usepackage[normalem]{ulem}
\usepackage[usenames,dvipsnames]{xcolor}
\usepackage{multirow}
\usepackage{mathtools}
\usepackage{placeins}
\usepackage[T1]{fontenc}
\usepackage[utf8]{inputenc}
\usepackage{hyperref}

\usepackage{nameref}
\usepackage{varioref}
\usepackage{hyperref}
\usepackage{cleveref}
\usepackage{comment}
\usepackage{graphicx}
\definecolor{orange}{rgb}{1,0.5,0}

\newcommand{\proj}[1]{\ket{#1}\!\bra{#1}}

\newcommand{\Tr}{\mathrm{Tr}}
\newcommand{\tr}{\mathrm{Tr}}

\begin{document}

\title{Optimising multipartite entanglement manipulation in quantum networks}

\author{Shashank Shekhar}
\affiliation{Department of Physics, Indian Institute of Technology Bombay, Mumbai, Maharashtra 400076, India}
\affiliation{Networked Quantum Devices Unit, Okinawa Institute of Science and Technology Graduate University, Okinawa, Japan}

\author{David Elkouss}
\affiliation{Networked Quantum Devices Unit, Okinawa Institute of Science and Technology Graduate University, Okinawa, Japan}
\affiliation{
QuTech, Delft University of Technology, Lorentzweg 1, 2628 CJ Delft, The Netherlands}

\author{Siddhartha Santra}
\email{Corresponding author: santra@iitb.ac.in}
\affiliation{Department of Physics, Indian Institute of Technology Bombay, Mumbai, Maharashtra 400076, India}
\affiliation{Center of Excellence in Quantum
Information, Computation, Science and Technology,
Indian Institute of Technology Bombay, Mumbai, Maharashtra 400076, India}

\begin{abstract}

Long-range, genuine multipartite entanglement (GME) between distant parties in a quantum network is a crucial resource for distributed quantum information processing. Through controlled manipulation of short-range GME states in the network via stochastic local operations aided by classical communication (SLOCC), long-range GME can be established between selected target nodes on demand. Optimizing such multipartite entanglement distribution protocols under constraints imposed by the network architecture, target-node set, and available resource budget is a key challenge in resource-efficient manipulation of GME. We present a general operational framework for manipulating GME in quantum networks by classifying the relevant networking operations into three broad primitives—inflation, deflation, and purification—reflecting their effect on the proliferation of GME through general bounds and illustrative examples. We show how these primitives can be combined to construct a broad class of multipartite quantum repeater protocols, enabling their systematic optimization, and demonstrate the framework with a simple example. Applying our approach to networks of Dür–Cirac–Tarrach states reveals a rich landscape of achievable end-to-end GME and accessible entanglement classes. More generally, our framework provides a systematic approach to the construction and optimization of multipartite quantum repeater protocols for scalable and resource-efficient GME distribution in quantum networks.

\end{abstract}

%\pacs{03.67.Hk, 03.67.Bg, 03.67.Pp}
\maketitle

\section{Introduction}
GME \cite{Horodecki2009RMP} is a property of multipartite quantum states in which intricate quantum correlations are shared among multiple ($N\geq 3$) parties in a manner that precludes their description as a mixture of partially separable states. Quantum states with GME exhibit a rich phenomenology, admitting inequivalent types of quantum correlations \cite{3qubitourestateclassicfication} and complex correlation patterns \cite{Acin2001,Siewert2012,Eltschka2012}. Operationally, such states are a critical resource for inherently multi-user protocols such as quantum secret sharing~\cite{Mark1999QSSintro}, clock synchronization~\cite{Komar2014CSintro}, $N$-party quantum key distribution~\cite{Epping2017NQKD}, and anonymous conference key agreement~\cite{Hah2020AQCKAintro}. %Harnessing GME as a scalable quantum network resource requires protocols to reliably distribute multipartite quantum entangled states between distant network nodes mitigating channel loss and noise while maintaining high-fidelity and at useful distribution rates.

Long-range distribution of GME in a quantum network can follow two broad approaches, both based on manipulating short-range entangled states to generate multipartite states between distant nodes. In the first approach \cite{2023WehnerMPEdistribution, Bugalho2023MPEdistribution}, bipartite entanglement is first established between selected node pairs using quantum repeaters \cite{briegel1998QRintro, Wehner2018Qnintro, Azuma2023QRintro}. The resulting long-range bipartite-entangled states are then converted into the target multipartite entangled state via SLOCC operations at the network nodes \cite{2023WehnerMPEdistribution}. The required operations—entanglement swapping and purification—are theoretically well understood \cite{swapping_santra, Santra2021swap, sohel_endtoend, bennett1EP, Deutsch1EP}, enabling the GME distribution rate to be expressed in terms of bipartite entanglement distribution rates \cite{Epping2017NQKD}.

Our focus here is on the second approach, in which the elementary short-range entangled resources are themselves multipartite states that are directly manipulated through SLOCC operations to establish long-range GME. Beyond its conceptual appeal, this approach can reduce qubit overhead, tolerate higher error thresholds, and achieve GME distribution rates comparable to the bipartite-based approach in certain scenarios \cite{Dur20162DQRintro, Dur20192DQRinto, Miguel20232DQRintro}.
This approach becomes necessary in networks whose elementary resource is itself multipartite. Such architectures offer a promising route toward large-scale quantum networks \cite{epping2016large, pirker2019quantum}, a prospect supported by recent proof-of-concept demonstrations \cite{multipartite_exp2}.

The multipartite entanglement manipulation approach requires a systematic understanding of the principal networking operations and their impact on GME as network states are processed to meet communication demands. Such a characterization is considerably more challenging than in the bipartite setting due to the complex structure of genuine multipartite entanglement. In particular, the absence of a multipartite analogue of the Schmidt decomposition precludes a simple description of entanglement evolution, while the freedom to choose different qubits for networking operations can lead to markedly different post-operation GME. Consequently, identifying resource-efficient manipulation protocols for a given network architecture and target-node set remains a challenging optimization problem, and existing multipartite quantum repeater schemes remain largely restricted to specific network topologies \cite{Dur20162DQRintro,Miguel20232DQRintro}, fixed swapping orders \cite{Dur20192DQRinto}, effectively bipartite elementary units \cite{2023WehnerMPEdistribution, Bugalho2023MPEdistribution, Wallnofer2019MSG}, or simple classes of multipartite resource states \cite{Dur20162DQRintro, Miguel20232DQRintro}. Recent experimental advances in multipartite entanglement manipulation \cite{mult_swap_exp,multipartite_expdetect} further motivate the development of a general operational framework for systematically designing and optimizing multipartite quantum repeater protocols under network and resource constraints.

In this work, we develop a general operational framework for manipulating genuine multipartite entanglement in quantum networks by organizing multipartite networking operations into three basic primitives—inflation, deflation, and purification—and analyzing their impact on genuine multipartite concurrence (GMC). We show how these primitives can be concatenated into multipartite quantum repeater protocols and demonstrate the framework on networks of Dür–Cirac–Tarrach (DCT) states, revealing a rich set of achievable end-to-end entanglement phases, including protocol-dependent transitions among separable, biseparable, W-type, and GHZ-type classes. We also formulate the corresponding optimization problem and illustrate, through a simple example, how the primitive-based framework can be used to optimize multipartite quantum repeater protocols for a given task, network architecture, and allowed set of operations. 

Unlike prior multipartite repeater protocols, which are fixed constructions for specific topologies, states, or swapping orders, our framework casts protocol design as an optimization over a common set of primitives, enabling systematic comparison and search across the protocol space for a given network and resource budget.

 % Inflationary operations, such as entanglement swapping~\cite{swap1,swap2}, merging~\cite{merging_exp,merging}, and entangling unitaries~\cite{Ent_unitaries}, distribute correlations across more parties, with the resulting GMC bounded by that of the input states. Deflationary operations remove parties through local measurements, redistributing entanglement subject to bounds determined by localizable entanglement~\cite{LME2017}, while purification probabilistically enhances GMC without changing the number of parties~\cite{Murao1998MEP5}.

\section{Results}

We begin by defining the quantum network model and the genuine multipartite entanglement measure used throughout this work, followed by a discussion of the three networking primitives and their concatenation into multipartite quantum repeater (M-QR) protocols, which forms the basis of the optimization framework.

% \textcolor{Blue}{We begin by first defining the quantum network model and the genuine multipartite concurrence measure used throughout this work. We then describe the three primitive operations for multipartite quantum networking, inflation, deflation, and purification, and derive general bounds on the GMC of the multipartite output states under these operations. Subsequently, we describe the concatenation of these primitives into a wide class of multipartite quantum repeater (M-QR) protocols. Applying two such M-QR protocols, in a network comprised of Dür–Cirac–Tarrach states as elementary units, which result due to local noise on $N$-party GHZ states and are therefore potentially experimentally relevant, we show that the GMC of the distributed final multipartite entangled state can belong to different SLOCC-inequivalent entanglement classes. Finally, we present a framework for optimisation of M-QR protocols and demonstrate its application for a network of GHZ states.}

% We begin by formally defining the quantum network model and the genuine multipartite entanglement measure used throughout. We then introduce the three primitive operations — inflation, deflation, and purification — and derive general bounds on their effect on GMC. These primitives are subsequently composed into multipartite quantum repeater (M-QR) protocols, applied to networks of DCT states, and cast as an optimization problem, which we illustrate on a simple network topology.

\subsection{Quantum network of multipartite entangled units and entanglement measure}
\label{subsec:2A}
\begin{figure}
    \centering
    \includegraphics[width=\linewidth]{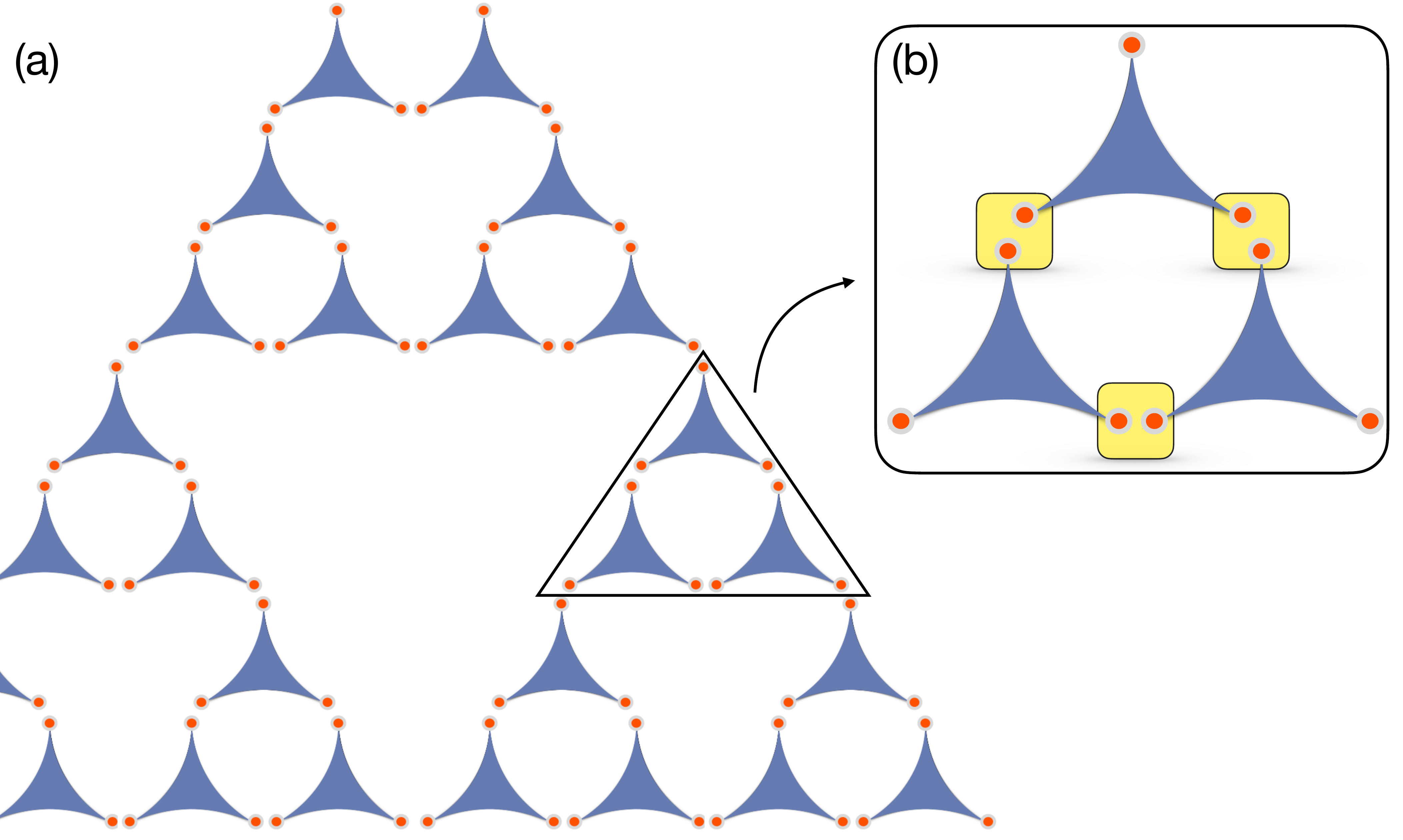}
    \caption{Schematic representation of a multipartite quantum network modeled as a hypergraph. (a) The network is composed of elementary three-qubit entangled units, with each blue triangular unit representing a hyperedge and the orange nodes representing qubits. (b) Enlarged view of the highlighted region in (a), illustrating how elementary hyperedges are connected through shared nodes to form a larger multipartite network.}
    \label{fig:triangle_net}
\end{figure}

We model the quantum network as a hypergraph $\mathcal{G}(V,\mathcal{E})$, where the vertices $V=\{v_i\}_{i=1}^M$ represent the network nodes and each hyperedge $e \in \mathcal{E}$ corresponds to short-range entangled units $\{\rho_{\mathcal N_i}\}_{i=1}^{N}$, where state $i$ contains $n_i$ qubits labeled by $Q_i$, with $|Q_i|=n_i$ and let \( \mathcal{Q}=\bigcup_{i=1}^{N} Q_i \) denote the set of all qubits in the network. By applying the networking operations detailed below, these elementary resource states can be manipulated to establish long-range multipartite entanglement between desired target nodes. For example, in Fig.~\ref{fig:triangle_net}, each small triangle represents an elementary multipartite entangled unit, or hyperedge, containing \(n_i=3\) qubits; these units can subsequently be combined through the networking operations described below to generate larger multipartite entangled states~\cite{Dur20162DQRintro}.

% Throughout, the parties relevant to genuine multipartite entanglement are the physical nodes, not individual qubits. Any operation carried out by a single node on qubits physically held in its own register is local with respect to the node partition, by the standard definition of locality in LOCC/SLOCC: locality is determined by the physical location performing the operation.

Throughout, all local operations are understood as SLOCC operations and may act jointly on any subset of qubits that are co-located at the same node. For the evaluation of genuine multipartite entanglement (GME), however, we treat all distinct qubits involved in the resulting state, after the specified operations have been performed, as distinct parties, irrespective of their physical location. Thus, co-location of qubits at a node does not by itself imply that they constitute a single party unless specified otherwise; rather, they are treated as a joint local register only when a joint operation acts on them.

As a measure of the multipartite entanglement we use the GMC, $C_g(\rho)$~\cite{Ma2011}. $C_g(\rho)$ defines the genuine multipartite entanglement in a mixed state, $\rho$, as the minimisation, 
\begin{align}
C_g(\rho)&=\inf_{\{p_i,|\psi_i\rangle\}}\sum_i p_i\,C_g(|\psi_i\rangle),
\end{align}
of the expected multipartite pure state linear-entropy based concurrence, $C_g(|\psi\rangle)=\min_{\gamma}\sqrt{2\big(1-\mathrm{Tr}\rho_{\gamma}^2\big)}$, over its possible pure state decompositions, $\rho=\sum_ip_i\ket{\psi_i}\bra{\psi_i}$.
For a pure multipartite state, $\ket{\psi}$, the GMC $C_g(\ket{\psi})$ equals the entanglement of the weakest bipartition and is strictly positive only if every bipartition is entangled. Throughout, we use GMC as our measure for GME; the primitive-based framework itself is measure-independent, though the quantitative bounds derived here (Eqs. \ref{eq:general_inflation_bound}, \ref{eq:general_deflation_bound}, \ref{eq:general_purification_bound}, \ref{eq:Cgmelb_rewritten}) are specific to GMC.

\subsection{Classes of multipartite quantum networking operations}
\label{subsec:2B}
Three principal types of operations can be used for multipartite entanglement manipulations and we describe them sequentially in the following.

\textbf{1. Inflation of multipartite entangled states:-} Inflation connects two multipartite states adjacent at a network node via measurements or unitary interactions between qubits from each state.

Consider a chain of $N$ short-range entangled units
$\{\rho_{\mathcal N_i}\}_{i=1}^N$, where the $i$th unit contains $n_i$
qubits labeled by $Q_i$, with $|Q_i|=n_i$. For each neighboring pair
$(i,i+1)$, let $a_i\in Q_i$ and $b_i\in Q_{i+1}$ denote the qubits on
which the inflation operation acts. If two elementary units also share
a node elsewhere in the network, the qubits that contribute to that
node are treated as separate unless a joint networking operation is performed on them. The initial state is $
\bigotimes_{i=1}^{N}\rho_{\mathcal N_i}.
$ For a sequence of inflation operations with outcome string
$\boldsymbol{\mu}=(\mu_1,\ldots,\mu_{N-1})$, we denote the corresponding
ordered composition by \(
    \hat O^{\boldsymbol{\mu}}_{\rm I}
    \equiv
    \hat O_{\rm I}^{\mu_{N-1}}
    \cdots
    \hat O_{\rm I}^{\mu_2}
    \hat O_{\rm I}^{\mu_1},
\)
where the rightmost operation acts first. Sequential (or, when
appropriate, parallel) inflation on the adjacent states is then
described by the stochastic map
\(
    \mathcal{E}_{\rm I}(\otimes_{i=1}^{N}\rho_{\mathcal N_i})
    =
    \left\{
        \left(P_{\boldsymbol{\mu}},
        \rho_{\boldsymbol{\mu}}^{R}\right)
    \right\}_{\boldsymbol{\mu}},
\)
where the conditional post-inflation state, supported on the residual
set $R$ of qubits, is
\begin{equation}
    \rho_{\boldsymbol{\mu}}^{R}
    =
    \frac{
        \Tr_{\left(\,\bigcup_{i=1}^N Q_i \setminus R\,\right)}
        \!\left[
            \hat O^{\boldsymbol{\mu}}_{\rm I}
            \left(\otimes_{i=1}^{N}\rho_{\mathcal N_i}\right)
            \hat O^{\boldsymbol{\mu} \dagger}_{\rm I}
        \right]
    }{
        \Tr\!\left[
            \hat O^{\boldsymbol{\mu}}_{\rm I}
            \left(\otimes_{i=1}^{N}\rho_{\mathcal N_i}\right)
            \hat O^{\boldsymbol{\mu} \dagger}_{\rm I}
        \right]
    },
    \label{eq:swap-squeezed}
\end{equation}
with outcome probabilities denoted by
\(
    P_{\boldsymbol{\mu}}.
\)
Here, $\hat O_{\rm I}^{\mu_i}$ specifies the inflationary operation
acting on the neighboring pair $(a_i,b_i)$ and, when measurements are
involved, is indexed by the corresponding measurement outcome $\mu_i$.

% For entanglement swapping implemented via a complete Bell-basis measurement, $\mu_i\in\{0,1,2,3\}$. For merging, realized by a CNOT between two qubits (one from each adjacent state) followed by a single-qubit measurement, $\mu_i\in\{0,1\}$. For entangling unitaries no measurement is performed, resulting in a single deterministic outcome (represented by fixing $\mu_i=0$ in our notation;

Denoting by, $\langle C_g^{\mathrm{(out)}}(\{\rho_{\mathcal N_i}\}) \rangle:=
\sum_{\boldsymbol{\mu}} P_{\boldsymbol{\mu}}\,C_g(\rho_{\boldsymbol{\mu}}^R)$, the expected output GMC of the ensemble of states obtained via inflationary operations, it can be  shown via the monotonicity of the GMC under SLOCC with respect to a suitable bipartition that,
\begin{equation}
\label{eq:general_inflation_bound}
\langle C_g^{\mathrm{(out)}}(\{\rho_{\mathcal N_i}\}) \rangle 
\le
\min_{1\le i\le N} C_g(\rho_{\mathcal N_i}),
\end{equation}
where, $C_g(\rho_{\mathcal N_i})$ are the GMCs of the input states.

\begin{proof}
Let the two multipartite pure states be
\[
\ket{\Psi}_{\mathcal A}=\ket{\psi}_{a_1\ldots a_n\,|\,A_1\ldots A_N},\qquad
\ket{\Phi}_{\mathcal B}=\ket{\phi}_{b_1\ldots b_m\,|\,B_1\ldots B_M},
\]
and denote their genuine–multipartite concurrences across the indicated bipartitions ({$a_i$} and {$b_i$} on one side and {$A_i$} and {$B_i$} on other) by
$C_g^{(\mathcal A)}$ and $C_g^{(\mathcal B)}$, respectively. Without loss of generality assume $C_g^{(\mathcal A)}\le C_g^{(\mathcal B)}$.  
Perform an inflation that acts only on qubits $a\in\mathcal A$ and $b\in\mathcal B$. One can always choose a global bipartition $X|X^c$ of the inflated system so that (i) the two sides of the original cut for $\mathcal A$ lie in opposite parts of $X|X^c$, and (ii) the local operations implementing the inflation act entirely on one side of $X|X^c$. By construction the initial entanglement across $X|X^c$ equals $C_g^{(\mathcal A)}$. Local operations and classical communication confined to one side of a bipartition cannot increase the entanglement across that bipartition on average. Hence, for the post-inflation ensemble $\{P_\mu,\rho_\mu\}_\mu$ we have
\begin{equation}
\langle C_g^{\mathrm{(out)}}\rangle=\sum_\mu P_\mu\,C_g(\rho_\mu)\le C_g^{(\mathcal A)}.
\label{eq:one_inflation_bound}
\end{equation}
Since $C_g^{(\mathcal A)}=\min\{C_g^{(\mathcal A)},C_g^{(\mathcal B)}\}$ by assumption, the bound \eqref{eq:general_inflation_bound} follows. Finally, the bound extends to arbitrary mixed inputs by the convex-roof construction: writing a mixed state as an ensemble of pure states and applying the above inequality to each pure component yields the same averaged upper bound for the mixed state. Iterating Eq.~(\ref{eq:one_inflation_bound}) bound for multiple iteration of inflation operations will lead to the upper bound of Eq.~(\ref{eq:general_inflation_bound}).

% The second part follows from the known results for entanglement swapping of bipartite entangled states, evaluated with respect to the bipartition between the measured qubit and the remaining subsystem~\cite{sohel_endtoend}.
\end{proof}

A second bound holds for a single entanglement-swapping step joining exactly two elementary units at one junction:
\(
\langle C_g^{(\mathrm{out})}\rangle
\leq
C_1^\ell C_2^\ell,
\)
where \(C_i^\ell\) is the bipartite concurrence across the connector qubit of unit \(i\). This follows because each unit has Schmidt rank \(\leq 2\) across its connector cut, regardless of its internal multipartite structure, so the four Bell outcomes combine exactly as in the bipartite case. This bound does not, in general, extend directly to a chain of successive entanglement-swapping operations, as illustrated by an example in the supplemental material.

% \begin{figure}
%     \centering
%     \includegraphics[width=\linewidth]{schematic all fig pdf.pdf}
%     \caption{(Color online) Schematic representation of multipartite networking operations and their effect on the entangled states in the network. Inflation connects neighboring entangled units increasing the number of parties involved in the entangled state and reduces the GMC whereas deflation disentangle selected parties reducing the number of qubits while possibly concentrating the GMC (not necessarily in general). Purification acts on multiple copies of a multipartite state and probabilistically enhances the GMC while preserving the number of parties involved.}
%     \label{fig:schematic}
% \end{figure}

% Ineq.~(\ref{eq:general_inflation_bound}) shows that multipartite entanglement distribution via swapping or merging in quantum networks is ultimately constrained by the bounds governing bipartite entanglement. Under repeated swapping or merging operations, the expected GMC can decay rapidly with the number of operations, $\prod_{j} C_j^\ell \sim e^{-(N-1)\ln C}$, thereby limiting the range over which multipartite correlations can be distributed \cite{mondal2024entanglement}. In contrast, when inflation proceeds solely through unitary interactions, Ineq.~(\ref{eq:general_inflation_bound}) implies that the expected GMC is limited only by the elementary unit with the weakest GME, suggesting greater robustness for distributing GME in the network via entangling unitaries.
Eq.~(\ref{eq:general_inflation_bound}) provides a general upper bound that holds for arbitrary families of states and inflationary operations. For certain families of states and specific networking operations, this bound can be further tightened to a multiplicative form that scales with the number of inflationary operations, $\langle C_g^{\mathrm{(out)}}(\{\rho_{\mathcal N_i}\})\rangle \leq \prod_i C_g(\rho_{\mathcal N_i})$. An important example is provided by Dür--Cirac--Tarrach (DCT) states under entanglement swapping, for which this multiplicative bound holds. Moreover, the bound is saturated for certain families of input states, including generalized GHZ states of the form $a\ket{0}^{\otimes n}+b\ket{1}^{\otimes n}$. In contrast, for generalized $W$-states, the expected output GMC can exceed the product of the input GMCs, demonstrating a qualitatively different scaling behavior from the GHZ family (see supplemental material for details).

% $a\ket{0}^{\otimes n}+b\ket{1}^{\otimes n}$ and the rank-two mixtures$
% \rho_{\text{mix}}
% =
% p |\Psi_{n}^{0+}\rangle\langle\Psi_{n}^{0+}|
% +
% (1-p)|\Psi_{n}^{0-}\rangle\langle\Psi_{n}^{0-}|,
% $,
% where,
% $
% \ket{\Psi_{n}^{0\pm}}
% =\tfrac{1}{\sqrt{2}}(\ket{0}^{\otimes n}\pm\ket{1}^{\otimes n}).
% $

\textbf{2. Deflation of multipartite entangled states :-} In deflation, qubits are projected or measured to disentangle them from the remaining state, reducing the multipartite state to fewer parties. Deflation operations  may be required when a private multipartite state is needed for a quantum task while preceding inflationary operations of short-ranged multipartite states results in a state on a larger number of parties.

For a multipartite state, $\rho$, on the set of parties, $Q=\{1,\dots,n\}$, let a subset of parties $r\subset Q$ be measured or projected out, with the residual state on the complement $R = Q \setminus r$ of size $(n-m)$. Then, the action of deflation operations can be described as a stochastic map, $
\mathcal{E}_{\rm D}(\rho)=\{(P_{\boldsymbol{\mu}},\rho_{\boldsymbol{\mu}}^R)\}_{\boldsymbol{\mu}}$, where $\boldsymbol{\mu}$ denotes the outcome, $P_{\boldsymbol{\mu}}$ the corresponding probability, and $\rho_{\boldsymbol{\mu}}^R$ the post-deflation state of the remaining parties given by,
\begin{align}
\rho_{\boldsymbol{\mu}}^R=\frac{\tr_{\mathbf r}[\hat O^{\boldsymbol{\mu}}_{\rm D}\rho(\hat O^{\boldsymbol{\mu}}_{\rm D})^\dagger]}{\tr[(\hat O^{\boldsymbol{\mu}}_{\rm D})^\dagger(\hat O^{\boldsymbol{\mu}}_{\rm D})\rho]},
\end{align}
where, $\hat O^{\boldsymbol{\mu}}_{\rm D}$ represents a deflationary operation on the set of $\mathbf r$ qubits. The simplest deflationary operations are independent projective measurements of the $m$ qubits, in which case $\hat O^{\boldsymbol{\mu}}_{\rm D}=\bigotimes_{i=1}^{m}\Pi^{\mu_i}$ with $\mu_i\in\{0,1\}$. This yields $2^m$ conditional states in the deflated ensemble corresponding to measurement outcome strings ${\mathbf \mu}\in\{0,1\}^{\otimes m}$. More generally, $\hat O^{\boldsymbol{\mu}}_{\rm D}$ may represent arbitrary local or collective quantum instruments, including adaptive and non-projective measurements which may not completely decouples the parties but only partially reduce their correlations with the rest of the network.

The expected output GMC for deflationary operations,
$
\langle C_g^{(\mathrm{out})}(\rho)\rangle
=
\sum_{\boldsymbol{\mu}} P_{\boldsymbol{\mu}}\, C_g(\rho_{\boldsymbol{\mu}}^R)$, is upper bounded by the localizable multipartite entanglement, $\mathscr{LE}(\rho)
=
\sup_{\hat O_{\rm D}}
\sum_{\boldsymbol{\mu}} P_{\boldsymbol{\mu}}\, C_g(\rho_{\boldsymbol{\mu}}^R)$, which optimizes over all possible measurement choices, $\hat{O}_{\rm D}=\{\hat O^{\boldsymbol{\mu}}_{\rm D}\}_{\boldsymbol \mu},\sum_{\boldsymbol \mu} (\hat O^{\boldsymbol{\mu}}_{\rm D})^\dagger \hat O^{\boldsymbol{\mu}}_{\rm D} =\mathbf{I}$, of the $m$-qubits ~\cite{LME2017}.
Thus, for any fixed measurement choice,
\begin{equation}
\langle C_g^{(\mathrm{out)}}(\rho)\rangle \le \mathscr{LE}(\rho).
\label{eq:general_deflation_bound}
\end{equation}
However, the magnitude of $\mathscr{LE}(\rho)$ relative to $C_g(\rho)$ depends on the pre-deflation state $\rho$, the choice of the measured qubits and measurement basis. Therefore, deflation, even with the optimal choice of local measurements, does not necessarily have a definite effect on the GME of network states and depending on the state family, measurements may reduce, preserve, or concentrate GME. 
 
For example, the three-qubit GHZ state, $
\left( |000\rangle + |111\rangle \right)/\sqrt{2}
$,
satisfies, $\mathscr{LE}(\rho) = C_{\mathrm{g}}(\rho)$ using Hadamard basis measurement of any single qubit, whereas the three-qubit W state $
\left( |001\rangle + |010\rangle + |100\rangle \right)/\sqrt{3}$
yields, $\mathscr{LE}(\rho) < C_{\mathrm{g}}(\rho)$, regardless of the basis in which anyone of the qubits is measured. Physically, this behavior reflects the fact that deflation can redistribute multipartite correlations among the remaining parties differently depending on the structure of the multipartite state: while some states such as GHZ retain their genuine multipartite entanglement structure under optimal measurements, others such as W states lose part of their multipartite entanglement, whereas in certain structured families like the DCT states, measurements can concentrate entanglement among the unmeasured parties.

% : while some states such as GHZ retain their genuine multipartite entanglement structure under optimal measurements, others such as W states lose part of their multipartite entanglement, whereas in certain structured families like the DCT states, measurements can concentrate entanglement among the unmeasured parties.

\textbf{3. Purification of multipartite entangled states:-} Multipartite entanglement purification (MEP) operation takes as input multiple noisy copies of multipartite entangled states and, via SLOCC, probabilistically outputs fewer copies with higher fidelity ~\cite{bennett1EP,bennett2EP,briegel1EP,durbriegel1EP,Zhou2020MEP,KrastanovMEP,JMiguel2024MEP}. The output of this operation is post-selected based on the outcome of the local operations at the network nodes.

MEP operations can be categorized into two classes — bipartite-based purification and multipartite-based purification~\cite{dur2003MEP3,dur2005MEP4,Murao1998MEP5,Huber2011MEP6} — distinguished by whether the purification procedure is constructed from bipartite entanglement purification primitives or employs genuinely multipartite primitives, such as P1--P2 method \cite{Murao1998MEP5} (See supplemental materials for details) or multi-subgraph schemes based on heterogeneous multipartite purification strategies~\cite{KrastanovMEP}.

A single multipartite purification step acting on \(m\) input copies \(\bigotimes_{k=1}^m \rho^{(k)}\) is described by the stochastic map, \(\mathcal{E}_{\rm P}(\bigotimes_{k=1}^m \rho^{(k)})=\{(P_{\boldsymbol{\nu}},\rho^{R}_{\boldsymbol{\nu}}\}_{\boldsymbol{\nu}}\), where, the string, \(\boldsymbol{\nu}=(\nu_1,\dots,\nu_\ell)\), denotes the outcome of applying the operators \(\hat{O}^{\boldsymbol{\nu}}_{\rm P}\) involved in the MEP operation (composed of local Cliffords/CNOTs, measurements and conditional unitaries, possibly acting with ancilla). The normalized postselected residual state is given by, 
\begin{equation}
\rho^{R}_{\boldsymbol{\nu}}
=\frac{\tr_{\rm anc}\!\Big[\,\hat{O}^{\boldsymbol{\nu}}_{\rm P}\,\big(\bigotimes_{k=1}^m \rho^{(k)}\big)\,{\hat{O}^{\boldsymbol{\nu}\dagger}_{\rm P}}\Big]}{P_{\boldsymbol{\nu}}},
\end{equation}
where `anc' denotes subsystems consumed or measured during the operation and $P_{\boldsymbol{\nu}}$ is the outcome probability. An iteration of the primitive is declared successful if $\boldsymbol{\nu}$ belongs to the success set
\(\mathcal{S}_{\rm succ}
\equiv
\left\{
\boldsymbol{\nu}:
C_g(\rho^{R}_{\boldsymbol{\nu}})\ge C_g(\rho)
\right\}.\) For every successful outcome \(\boldsymbol{\nu}\in\mathcal{S}_{\rm succ}\), the residual GMC satisfies
\begin{equation}
C_g(\rho)
\le
C_g(\rho^{R}_{\boldsymbol{\nu}})
\le
\frac{C_g\left(\bigotimes_{k=1}^{m}\rho^{(k)}\right)}
{P_{\boldsymbol{\nu}}},
\label{eq:general_purification_bound}
\end{equation}
where for the upper bound, the \(m\)-copy input state is regarded as an \(N\)-partite state by grouping the \(m\) subsystems associated with each physical node into a single local party. Thus, the relevant partitions for $C_g$ here becomes the partitions of the physical nodes, rather than of the individual qubits.

% Each of the $m$ input copies contributes one qubit to node local register; all operations thus act within individual nodes' registers and are therefore local.

% For every successful outcome $\boldsymbol{\nu}\in\mathcal{S}_{\rm succ}$, the residual GMC therefore satisfies
% \begin{equation}
% C_g(\rho)
% \;\le\;
% C_g(\rho^{R}_{\boldsymbol{\nu}})
% \;\le\;
% \frac{m\,C_g(\rho)}{P_{\boldsymbol{\nu}}},
% \label{eq:general_purification_bound}
% \end{equation}
% with overall success probability \(P_{\rm succ}
% =
% \sum_{\boldsymbol{\nu}\in\mathcal{S}_{\rm succ}}
% P_{\boldsymbol{\nu}}.\)

% The lower bound follows directly from the definition of $\mathcal{S}_{\rm succ}$: an outcome is accepted only if the resulting residual state has GMC at least as large as that of a single input copy.

% For the upper bound, we use two properties of $C_g$, namely monotonicity under LOCC instruments and subadditivity under tensor products (discussed in \cite{GMC_prop}) of copies held by the same parties which gives
% \begin{equation}
% \sum_{\boldsymbol{\nu}}
% P_{\boldsymbol{\nu}}\,
% C_g(\rho^{R}_{\boldsymbol{\nu}})
% \le
% C_g\!\left(
% \bigotimes_{k=1}^{m}\rho^{(k)}
% \right) \le
% m\,C_g(\rho),
% \end{equation}
% where $\nu$ runs over the whole set of outcomes and therefore, for every $\nu \in \mathcal{S}_{\rm succ} $ gives the upper bound and hence Eq.~\eqref{eq:general_purification_bound}.

% Purification may be required for long-range entanglement distribution, as it mitigates the degradation in GMC induced by successive inflation and deflation operations and increases the attainable GMC.

\subsection{Multipartite entanglement–based quantum repeater (M-QR) protocols by concatenating the primitive operations}
\label{subsec:2C}
 A multipartite quantum repeater (M-QR) protocol can be defined as a finite composition of the primitive networking operations, $\mathcal{E}_{\rm I}$ (inflation), $\mathcal{E}_{\rm D}$ (deflation), and $\mathcal{E}_{\rm P}$ (purification), acting on a collection of short-range entangled units $\{\rho_{\mathcal N_i}\}$ in a quantum network---in direct analogy with first-generation quantum repeaters~\cite{Muralidharan2016} for bipartite entanglement distribution. The overall M-QR protocol may be schematically expressed as
\begin{equation}
\mathcal{E}_{\rm M\text{-}QR}
\!\left(
\bigotimes_{i=1}^{N}
\rho_{\mathcal N_i}
\right)
=
\mathcal{E}_{\alpha_k}^{(\mathcal S_k)}
\circ
\cdots
\circ
\mathcal{E}_{\alpha_2}^{(\mathcal S_2)}
\circ
\mathcal{E}_{\alpha_1}^{(\mathcal S_1)}
\!\left(
\bigotimes_{i=1}^{N}
\rho_{\mathcal N_i}
\right),
\label{eq:M-QR_map}
\end{equation}
where $\alpha_\ell \in \{\mathrm I,\mathrm D,\mathrm P\}$ specifies the primitive employed at step $\ell$, and $\mathcal S_\ell \subseteq \mathcal Q$ denotes the subset of network qubits on which that primitive acts, with $\mathcal Q$ the set of all qubits in the network. In general, the choice of a primitive and its support $\mathcal S_\ell$ at a given step may depend adaptively on the classical outcomes generated by one or more preceding operations. Each primitive produces a post-operation state conditioned on the corresponding classical outcome string. Existing protocols fit naturally within this framework. In particular, we recast the 2D M-QR protocols for GHZ, W and graph states of Refs.~\cite{Dur20162DQRintro,Miguel20232DQRintro,HahnPappaEisert2019} in terms of our inflation, deflation, and purification primitives.

For the GHZ protocol~\cite{Dur20162DQRintro}, each iteration consists of two inflation steps followed by a deflation step. The noise model considered is independent single-qubit depolarizing noise which brings the initial GHZ state to a DCT state of the form of Eq.~(\ref{eq:elementaryunits}). The two inflation steps are implemented by Bell-state measurements (BSMs) that fuse three neighboring tripartite DCT units into a five-party state. Since DCT states are diagonal in the GHZ basis and different BSM outcomes are related by local Pauli corrections, all $16$ possible outcome strings of the two inflation steps are LOCC-equivalent. Hence, the state entering the deflation step, and in particular its GMC, is independent of the inflation outcomes.

The deflation step is implemented by a third BSM, which removes two qubits and produces a three-party output state $\rho^{\rm out}$. In contrast to the inflation stage, the four possible deflation outcomes are not all LOCC-equivalent. They can therefore lead to different output branches, with some satisfying $C_g(\rho^{\rm out})>0$ and others satisfying $C_g(\rho^{\rm out})=0$. Post-selecting the branches with nonzero GMC consequently yields a larger GMC than the outcome-averaged value.

Purification $\mathcal P$ may be inserted between successive iterations when required, depending on whether the resulting state remains above the relevant distillability threshold. Thus, a single iteration of the GHZ protocol is represented by
\begin{equation}
\mathcal{E}_{\rm M\text{-}QR}^{\rm GHZ}
:=
\mathcal{E}_{\mathrm D}^{\rm BSM}
\circ
\mathcal{E}_{\mathrm I}^{\rm BSM}
\circ
\mathcal{E}_{\mathrm I}^{\rm BSM}.
\end{equation}

The W-state protocol~\cite{Miguel20232DQRintro} has the same two-inflation--one-deflation architecture with an additional conditional deflation step. Each iteration begins with three noisy tripartite W states. The two inflation steps are implemented by parity measurements $P_e$ and $P_o$ on qubits belonging to different units, denoted by $\mathcal{E}_{\mathrm I}^{\rm Parity}$.

The deflation stage is initiated by a third parity measurement, $\mathcal{E}_{\mathrm D}^{\rm Parity}$, which projects the state onto an even or odd subspace. The subsequent operations depend on the outcomes of the preceding parity measurements. One branch is completed by Bell-state measurements (BSMs), while the other requires a computational-basis measurement conditioned on a specific outcome, followed by BSMs. We denote these conditional operations collectively by $\mathcal{E}_{\mathrm D}^{\rm BSM/Z}$. In either successful branch, the remaining three qubits are locally equivalent to a W state.

The protocol fails when all three parity outcomes coincide, resulting in a fixed success probability of $2/3$. Thus, a single iteration of the W-state protocol is represented by
\begin{equation}
\mathcal{E}_{\rm M\text{-}QR}^{\rm W}
:=
\mathcal{E}_{\mathrm D}^{\rm BSM/Z}
\circ
\mathcal{E}_{\mathrm D}^{\rm Parity}
\circ
\mathcal{E}_{\mathrm I}^{\rm Parity}
\circ
\mathcal{E}_{\mathrm I}^{\rm Parity}.
\end{equation}

The additional conditional step in the W-state protocol reflects the inequivalent entanglement structure of W and GHZ states.

% Explicit forms of the inflation and deflation operations, together with the corresponding outcome-dependent corrections, are given in the Supplemental Material.

In the graph state protocol \cite{HahnPappaEisert2019}, the routing operation takes an already shared graph state as its input. To establish an EPR pair between any two nodes, the protocol first selects a shortest path connecting them and measures the intermediate path qubits in the $X$ basis. These measurements remove the intermediate vertices while modifying the graph connectivity via local complementation. The remaining unwanted vertices in the neighborhoods of end nodes are then measured in the $Z$ basis, leaving an EPR pair between the two endpoints. Thus, the routing operation can be schematically decomposed as
\begin{equation}
\mathcal{E}_{\mathrm{M-QR}}^{\mathrm{graph}}
=
\mathcal{E}_{\mathrm D}^{Z}
\circ
\mathcal{E}_{\mathrm D}^{X},
\end{equation}
where $\mathcal{E}_{\mathrm D}^{X}$ is understood to include the local-complementation effects associated with the $X$ measurements. The same decomposition can be used to describe the GHZ-extraction examples considered in the protocol~\cite{HahnPappaEisert2019}.

We also emphasize that, although the GMC of the final state can, in principle, be computed exactly by tracking the evolution under Eq.~(\ref{eq:M-QR_map}), this quickly becomes intractable for large networks. For an $N_q$-qubit state, evaluating the GMC requires concurrence calculations over all $(2^{N_q-1}-1)$ bipartitions, and the result depends on the specific inflation and deflation measurements performed. Instead, for some special structured families, efficient estimates can be obtained by successively applying the bounds in Ineqs.~(\ref{eq:general_inflation_bound}), (\ref{eq:general_deflation_bound}), and (\ref{eq:general_purification_bound}) to the constituent primitives, enabling the analysis of large-scale M-QR protocols without exact state evolution. One such example of special family is D\"ur--Cirac--Tarrach (DCT) states discussed in the following section. 

% The GMC of the final multipartite state produced by an M-QR protocol can, in principle, be computed exactly by tracking the state evolution under the primitives defining the protocol, Eq.~(\ref{eq:M-QR_map}). However, this becomes computationally demanding as the network size grows: for a state of $N_q$ qubits, the evaluation of GMC requires $2^{N_q}$ bipartite concurrence calculations. Moreover, the exact value depends on the specific qubits chosen for inflationary or deflationary measurements. Instead, efficient estimates of the GMC can be obtained by applying the bounds in Ineqs.~(\ref{eq:general_inflation_bound}), (\ref{eq:general_deflation_bound}), and (\ref{eq:general_purification_bound}) sequentially to the constituent primitives to obtain estimates of the final GMC, thereby enabling the study of M-QR protocols in settings where exact simulations become impractical. 

% The M-QR framework extends naturally to larger multipartite networks which can exhibit qualitatively richer behavior. For example, in a four-qubit square architecture similar to triangular architecture in \cite{Dur20162DQRintro}, post-selected multi-qubit deflation can simultaneously expand the distributed state and enhance the GMC beyond that of the elementary units an effect absent in the three-qubit protocol. This effect may substantially reduce the resource overhead of MEP and can be particularly pronounced for certain input states. See supplemental material for details.

\subsection{Multipartite networking with DCT states}
\label{subsec:2D}DCT states~\cite{1999DCTintro} provide a natural family of elementary multipartite resource states for quantum networks. They are the multipartite analogue of Werner states. More generally, any $N$-qubit state can be transformed by LOCC into a DCT state through a depolarization (twirling) procedure that preserves its overlap with the target GHZ state. Thus, for GHZ-based network protocols, an arbitrary noisy multipartite resource can be associated with a canonical DCT representative having the same GHZ overlap, providing a natural and physically meaningful model for the elementary resources considered here. These states can also arise from local depolarizing noise acting on an $N$-qubit Greenberger-Horne-Zeiliner (GHZ) state and take the form,
\begin{equation}
\rho(\lambda_c,\{\lambda_i\})
= \lambda_c\,\proj{\Psi_n^{0+}}
+ \sum_{j=0}^{2^{\,n-1}-1}\lambda_j \sum_{\sigma=\pm}\proj{\Psi_n^{j\sigma}},
\label{eq:elementaryunits}
\end{equation}
where \(\ket{\Psi_n^{j\pm}}=\tfrac{1}{\sqrt{2}}\bigl(\ket{0\,j}\pm\ket{1\,\overline{j}}\bigr)\), \(j\) is an \((n-1)\)-bit string, \(\overline{j}\) its bitwise complement, and the weights satisfy \(\lambda_c~\!,\lambda_j\ge0\) and are normalised, \(\lambda_c+2\sum_j\lambda_j=1\)~\cite{Huber2012}. The GMC for such states is linearly related to their fidelity relative to the GHZ state ~\cite{Huber2012}, $
C_g(\rho)={\max}\{0,2\bigl(\lambda_c+\lambda_0\bigr)-1\}={\max}\{0,2\braket{\Psi_{n}^{0+}|\rho|\Psi_{n}^{0+}}-1\}$. 

\begin{figure}
    \centering
    \includegraphics[width=\linewidth]{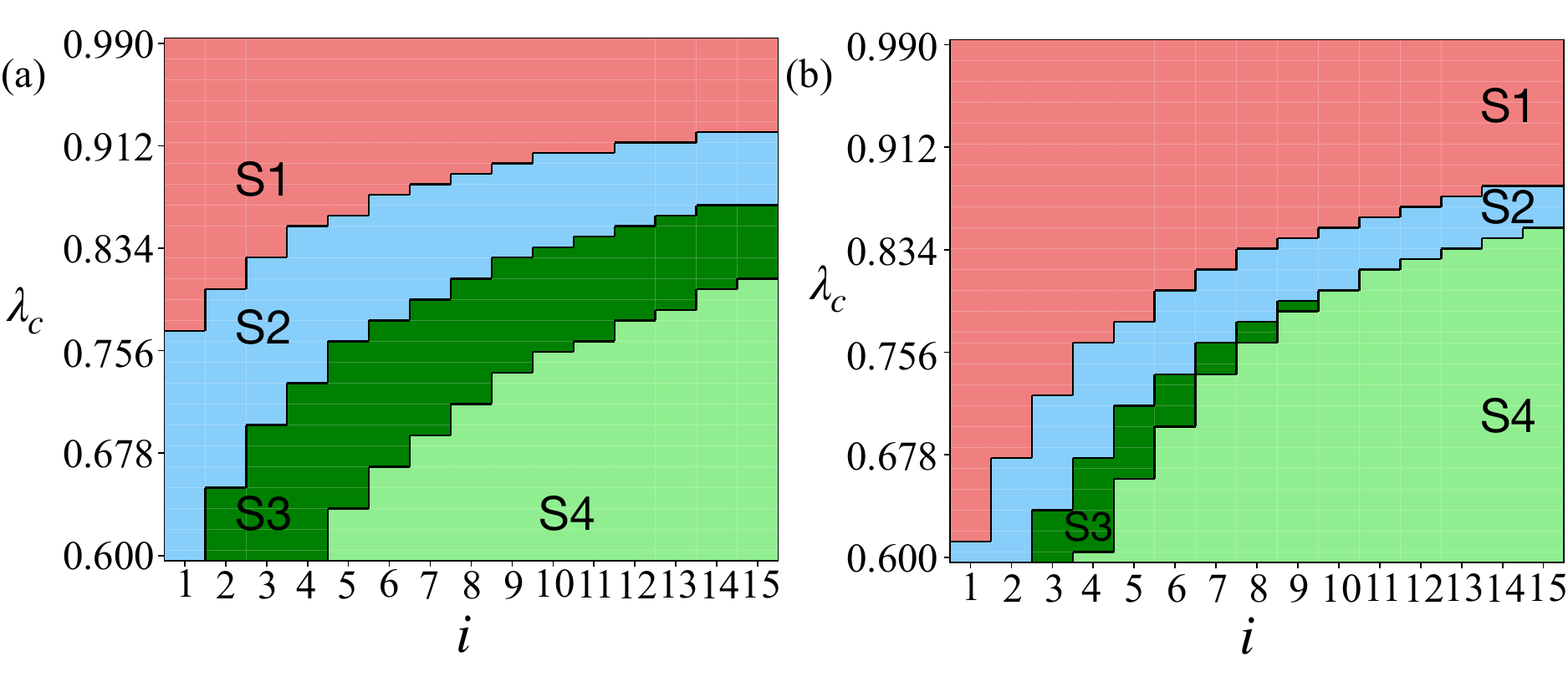}
     \caption{Multipartite entanglement class of the final DCT state as a function of PCM iterations ($i$) of the M-QR protocol for different initial parameters of tripartite DCT states. (a) PCM protocol. (b) PCM protocol with end-to-end purification. The regions correspond to different entanglement classes: red ($S_1$) GHZ-class states identified by a nonzero three-tangle; blue ($S_2$) genuinely multipartite entangled states that may belong to either the GHZ or W class; green ($S_3$) biseparable states; light green ($S_4$) fully separable states.}

    \label{fig:Cgme evolution tripartite 4}
\end{figure}

Importantly, DCT states remain within the DCT family under swapping and merging. This property, together with Eq.~\ref{eq:general_inflation_bound}, implies an upper bound that scales multiplicatively with the number of swapping operations. In particular, $N$ elementary units, $\{\rho_{\mathcal N_i}\}_{i=1}^N$, of the form of Eq.~\ref{eq:elementaryunits}, with GMCs, \(C_g^{(i)}:=C_g(\rho_{\mathcal N_i})\), when sequentially inflated by entanglement swapping, as described in Eq.~(\ref{eq:swap-squeezed}) yield a DCT state on $(\sum_i n_i-2(N-1))$ qubits with expected GMC that satisfies,

\begin{equation}
\label{eq:Cgmelb_rewritten}
\max\!\left\{
0,\,
\frac{\prod_{i=1}^{N}\!\left(1+C_g^{(i)}\right)}
     {2^{N-1}}-1
\right\}
\;\le\;
\big\langle C_g^{\mathrm{(out)}}\big\rangle
\;\le\;
\prod_{i=1}^{N} C_g^{(i)} 
\end{equation}
% \begin{equation}
% \label{eq:Cgmelb_rewritten}
% \max\!\left\{
% 0,\,
% 2^{1-N}\!\prod_{i=1}^{N}\!\left(1+C_g^{(i)}\right)-1
% \right\}
% \;\le\;
% \big\langle C_g^{\mathrm{(out)}}(\{\rho_{\mathcal N_i}\})\big\rangle
% \;\le\;
% \prod_{i=1}^{N} C_g^{(i)} .
% \end{equation}

% \begin{equation}
% \label{eq:Cgmelb_rewritten}
% \max\Bigl(0,\frac{\displaystyle\prod_{i=1}^{N}\big(C_g^{(i)}+1\big)}{2^{\,N-1}}-1\Bigr) \;\le\; 
% \langle C_g^{\mathrm{(out)}}(\{\rho_{\mathcal N_i}\}) \rangle
% \;\le\;
% \prod_{i=1}^{N} C_g^{(i)}
% ,
% \end{equation}
independent of the swapping order or choice of qubits measured where $\big\langle C_g^{\mathrm{(out)}}\big\rangle \equiv \big\langle C_g^{\mathrm{(out)}}(\{\rho_{\mathcal N_i}\})\big\rangle$. Further, for DCT states, the optimal single-qubit orthogonal projective measurement is the Hadamard-basis measurement which gives $\langle C_g^{\mathrm{(out)}}(\rho) \rangle  \geq C_{\mathrm{g}}(\rho)$ (see supplemental material for the proofs).

To illustrate Eq.~(\ref{eq:M-QR_map}), we onsider a linear chain of tripartite DCT elementary units $\rho(\lambda_c,{\lambda_i})$ of the form in Eq.~(\ref{eq:elementaryunits}), and apply a M-QR protocol iteratively to distribute a long-range tripartite state. Each iteration of the protocol consists of an inflation step followed by a deflation step, with the number of qubits preserved after every round. We refer to this as the party-conserving mode (PCM) of the M-QR protocol.

The resulting state can belong to different class of the tripartite entanglement, and successive PCM rounds can drive transitions between these regions. For the three-qubit DCT family considered here, we distinguish the fully separable, biseparable, W-class, and GHZ-class regions using the genuine multipartite concurrence $C_g$, the three-tangle $\tau_3$, and the PPT criterion. States in the GHZ class have $\tau_3>0$, while W-class states have $\tau_3=0$ and $C_g>0$. The remaining states with $C_g=0$ are separated into fully separable and biseparable states by their bipartite PPT properties: fully separable states are PPT across every bipartition, whereas, within the DCT family considered here, biseparable states are non-PPT across at least one bipartition and satisfy $\tau_3=0$ \cite{1999DCTintro,Eltschka2012,Siewert2012}.Thus, $C_g$ first distinguishes the genuinely multipartite-entangled
sector from the $C_g=0$ sector, while $\tau_3$ and the PPT criteria
resolve the entanglement classes within these regions.

For the present family, however, only a closed-form lower bound on $\tau_3$ is available \cite{Eltschka2012,Siewert2012}. Hence, this bound alone does not allow the GHZ and W regions to be distinguished exactly. Accordingly, the region labelled $S_2$ in the left panel of Fig.~\ref{fig:Cgme evolution tripartite 4} should be understood as states that may belong to either the GHZ or W class.

We further investigate how these regions are modified when purification is interleaved with the PCM protocol. Specifically, an end-to-end purification step is applied after each PCM round, thereby altering the state evolution and, consequently, the extent of the different entanglement regions. The resulting phase structure after four purification rounds is shown in the right panel of Fig.~\ref{fig:Cgme evolution tripartite 4}.

\subsection{Optimising M-QR protocols}
\label{subsec:2E}
M-QR protocols synthesized from the elementary primitives, as defined in Eq.~(\ref{eq:M-QR_map}), can in principle be systematically optimized over the full space of primitive choices, qubit supports, and measurement settings. This enables exploration of the broader protocol space rather than restricting the analysis to ad hoc constructions. The motivation for such an optimization is illustrated in the following section through a simple example, by showing that, even for a fixed network topology, different choices of inflation and deflation operations can lead to substantially different values of the metrics of interest, highlighting the importance of systematic protocol optimization.

These observations highlight the importance of an optimization framework for M-QR design. The optimization variables would comprise of the sequence of elementary primitives and the subsets of qubits on which they act, while the network architecture, target-node set, and available resource budget define the constraints. Depending on the application, the objective may be to maximize the final GMC, success probability, or resource efficiency, such as by minimizing the number of gate operations, measurements, or qubits required. The resulting protocol optimization problem may be expressed as
\begin{equation}
\{(\alpha_\ell^\star,\mathcal S_\ell^\star)\}
=
\arg\max_{\{(\alpha_\ell,\mathcal S_\ell)\}}
\mathcal{F}
\!\left[
\mathcal{E}_{\alpha_k}^{(\mathcal S_k)}
\circ\cdots\circ
\mathcal{E}_{\alpha_1}^{(\mathcal S_1)}
\!\left(
\bigotimes_{i=1}^{N}\rho_{\mathcal N_i}
\right)
\right],
\label{eq:optimisation}
\end{equation}
where $\alpha_\ell\in\{\mathrm I,\mathrm D,\mathrm P\}$ specifies the primitive employed at step $\ell$, $\mathcal S_\ell$ denotes the subset of qubits on which it acts, and $\mathcal F$ is a real-valued performance metric evaluated on the output state of the M-QR protocol. A primitive based optimization approach has been very useful for linear quantum repeaters~\cite{liangjiang2007dynamicprogramming}, likely the primitive-based framework developed here can be exploited analogously for the systematic exploration and benchmarking of M-QR protocols.

% Successive rounds of PCM drive transitions between the different entanglement regions (left panel of Fig.~\ref{fig:Cgme evolution tripartite 4})--fully separable, biseparable, W-type, and GHZ-type — distinguished by the three-tangle $\tau_3$ (GHZ-type states have $\tau_3>0$; W-type states have $\tau_3=0$ but $C_{g}>0$). Since we only have a closed expression for lower bound for $\tau_3$, thus we cannot separate the GHZ and W class exactly giving the S2 region which may either be GHZ or W class. For DCT states, full separability coincides with positivity under all partial transpositions, while biseparable states satisfy $\tau_3=0$ and $C_{g}=0$ but are non-PPT in at least one bipartition \cite{1999DCTintro,Eltschka2012,Siewert2012}. Incorporating end-to-end purification after each PCM round further modifies the extent of these regions, as illustrated in the right panel for four purification rounds.

\subsection{Case study: M-QR protocol comparison on fixed topology}
\label{subsec:2F}
For a fixed network topology, multiple M-QR protocols can generate the same multipartite target state while yielding different values of the performance metric \(\mathcal{F}\), arising solely from differences in the choice and ordering of primitive operations. Consequently, protocol performance is not determined by topology alone: the optimal protocol may depend on the underlying noise model and the chosen figure of merit. Even for small network topologies, we find that the optimal protocol can vary across these conditions, motivating a systematic optimization over the space of admissible M-QR protocols.

Therefore, we instantiate the optimization problem of Eq. (\ref{eq:optimisation}) with a simple illustrative example on a fixed topology (Fig.~\ref{fig:triangle_net}b and ~\cite{Dur20162DQRintro}), evaluating the argmax under three different choices of performance metric $\mathcal{F}$ and noise model — average/postselected concurrence under deterministic noise, success probability under deterministic noise, and average concurrence under probabilistic link generation with noise applied to available qubits at every time step — to show explicitly how the optimal protocol depends on this choice. See the supplemental material for more such comparison.

% BSMs during inflation followed by a third BSM during deflation, reproducing the original triangular M-QR protocol of Ref.~\cite{Dur20162DQRintro}.
% uence of Ref.~\cite{Dur20162DQRintro} as the optimal protocol for every performance metric considered except the average fidelity. The only exception is $\langle F\rangle$, for which the hybrid protocol consisting of two BSM inflations followed by Hadamard-basis deflation performs better. This already demonstrates that even under deterministic operation the optimal M-QR strategy depends on the chosen figure of merit rather than solely on the underlying noise model.

\begin{figure}
    \centering
    \includegraphics[width=0.9\linewidth]{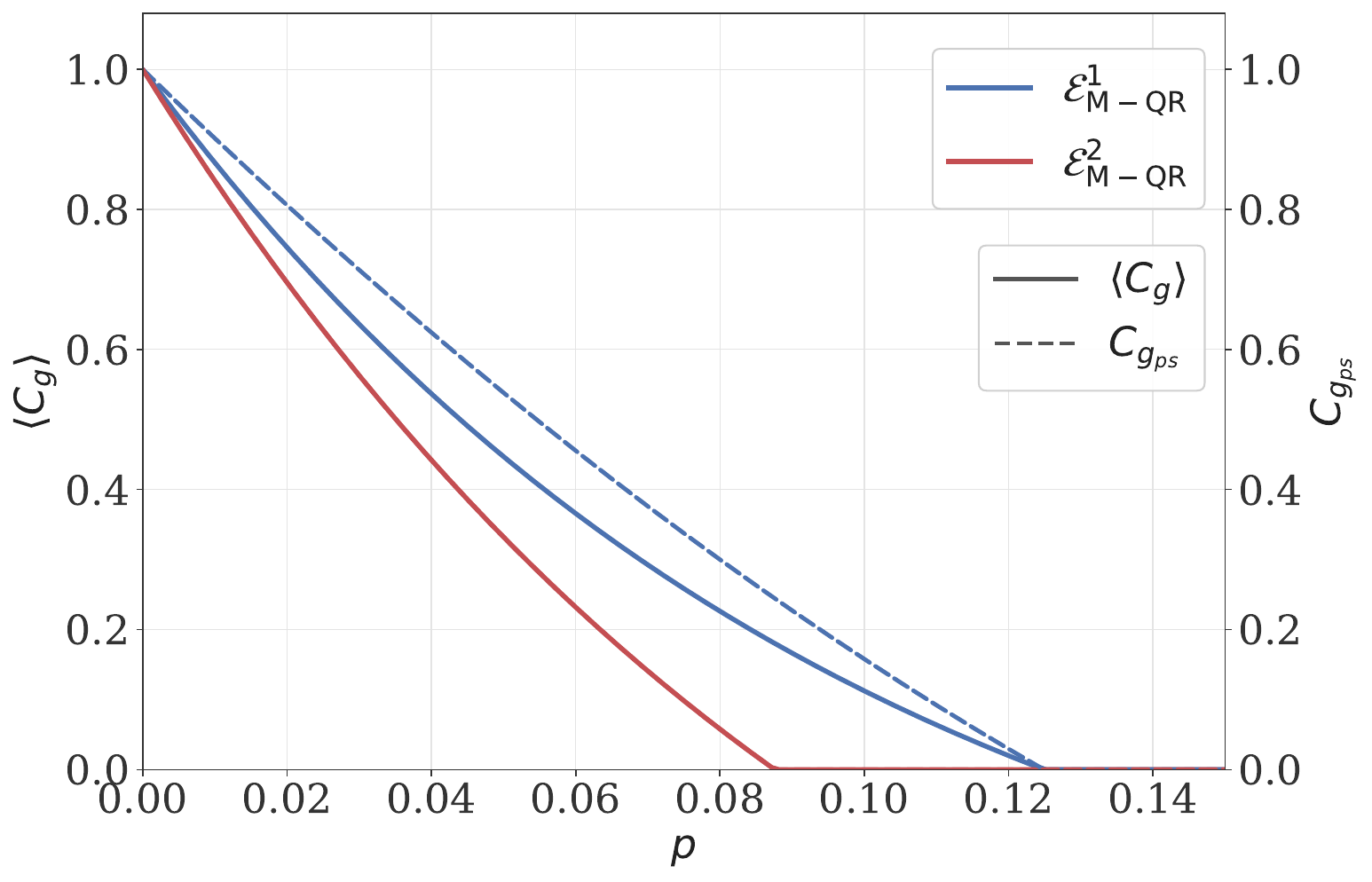}
    \caption{Average ($\langle C_g\rangle$, solid) and best postselected ($C_{g_{\rm ps}}$, dashed) genuine multipartite concurrence versus SQDN strength $p$, comparing the all-BSM protocol $\mathcal{E}^1_{\rm M-QR}$ (blue) and the hybrid BSM+Hadamard protocol $\mathcal{E}^2_{\rm M-QR}$ (red). $\mathcal{E}^1_{\rm M-QR}$ dominates both metrics over the full range of $p$.}
    \label{fig:triangle_SQDN}
\end{figure}

\begin{figure}
    \centering
    \includegraphics[width=0.9\linewidth]{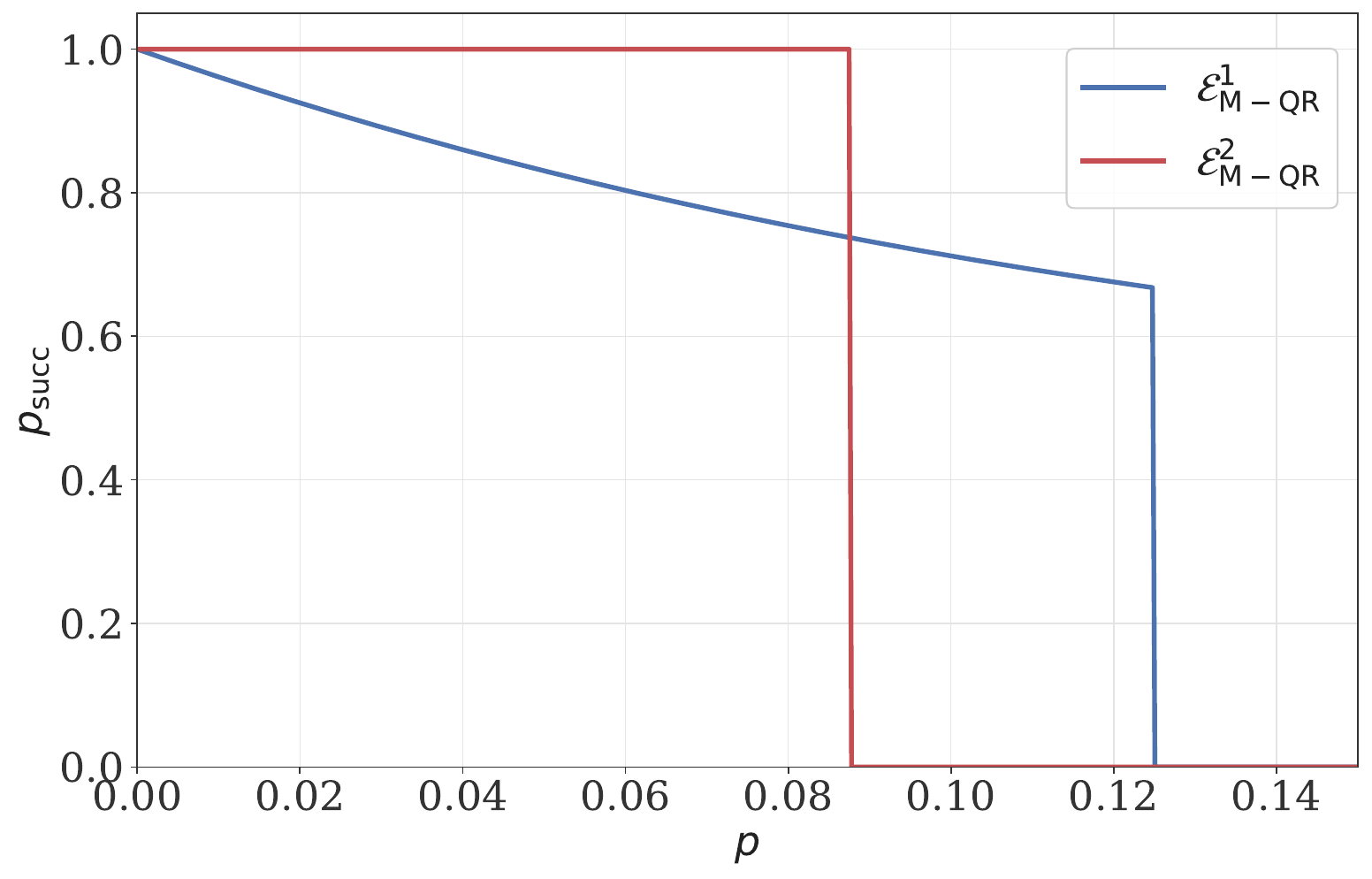}
    \caption{Success probability $p_{\mathrm{succ}}$ of the two triangular M-QR protocols as a function of SQDN strength $p$. $E^{2}_{\mathrm{M\text{-}QR}}$ retains $p_{\mathrm{succ}}=1$ up to $p\approx0.087$ before dropping sharply to zero, whereas $E^{1}_{\mathrm{M\text{-}QR}}$ declines gradually and remains nonzero up to $p\approx0.125$. The curves cross near $p\approx0.087$, above which $E^{1}_{\mathrm{M\text{-}QR}}$ has both higher $C_g$ and $p_{\mathrm{succ}}$.}
    \label{fig:triangle_p_succ}
\end{figure}

The first protocol, corresponding to the original scheme of Ref.~\cite{Dur20162DQRintro}, employs a third BSM for deflation and is given by
\begin{equation}
\mathcal{E}_{\mathrm{M\text{-}QR}}^{1}
:=
\mathcal{E}_{\mathrm{D}}^{\mathrm{BSM}}
\circ
\mathcal{E}_{\mathrm{I}}^{\mathrm{BSM}}
\circ
\mathcal{E}_{\mathrm{I}}^{\mathrm{BSM}}.
\end{equation}
The second protocol replaces the final BSM with two independent Hadamard-basis measurements on the auxiliary qubits, yielding
\begin{equation}
\mathcal{E}_{\mathrm{M\text{-}QR}}^{2}
:=
\mathcal{E}_{\mathrm{D}}^{H\otimes H}
\circ
\mathcal{E}_{\mathrm{I}}^{\mathrm{BSM}}
\circ
\mathcal{E}_{\mathrm{I}}^{\mathrm{BSM}},
\end{equation}
where $\mathcal{E}_{\mathrm{D}}^{H\otimes H}$ denotes the independent Hadamard-basis measurements of the two auxiliary qubits.

\paragraph*{Deterministic link generation.}
For independent single-qubit depolarizing noise (SQDN), we compare $\mathcal{E}^1_{\rm M-QR}$ and $\mathcal{E}^2_{\rm M-QR}$ using the average genuine multipartite concurrence $\langle C_g\rangle$ and the best postselected concurrence $C_{g_{\rm ps}}$. As shown in Fig.~\ref{fig:triangle_SQDN}, $\mathcal{E}^1_{\rm M-QR}$ yields higher value than $\mathcal{E}^2_{\rm M-QR}$ for both metrics across the full noise range, confirming the all-BSM protocol as optimal for this figure of merit under SQDN.

However, entanglement quality is not the only relevant resource for repeater operation: the success probability of each protocol---defined here as the probability that a given run yields a nonzero-GMC output branch rather than a discarded (zero-GMC) one---is itself an important figure of merit. Fig.~\ref{fig:triangle_p_succ} shows $p_{\mathrm{succ}}$ as a function of the SQDN strength $p$ for both protocols. $\mathcal{E}^{2}_{\mathrm{M\text{-}QR}}$ maintains $p_{\mathrm{succ}}=1$ up to $p\approx 0.087$, beyond which it falls sharply to zero, whereas $\mathcal{E}^{1}_{\mathrm{M\text{-}QR}}$ declines smoothly from $p_{\mathrm{succ}}=1$ and persists---with nonzero success probability---up to $p\approx 0.125$. The two curves cross near $p\approx 0.087$: for $p\lesssim 0.087$, $\mathcal{E}^{2}_{\mathrm{M\text{-}QR}}$ has the higher success probability despite its lower $C_g$ , while for $p\gtrsim 0.087$, $\mathcal{E}^{1}_{\mathrm{M\text{-}QR}}$ dominates in both $C_g$ and $p_{\mathrm{succ}}$ simultaneously.

\begin{figure}
    \centering
    \includegraphics[width=\linewidth]{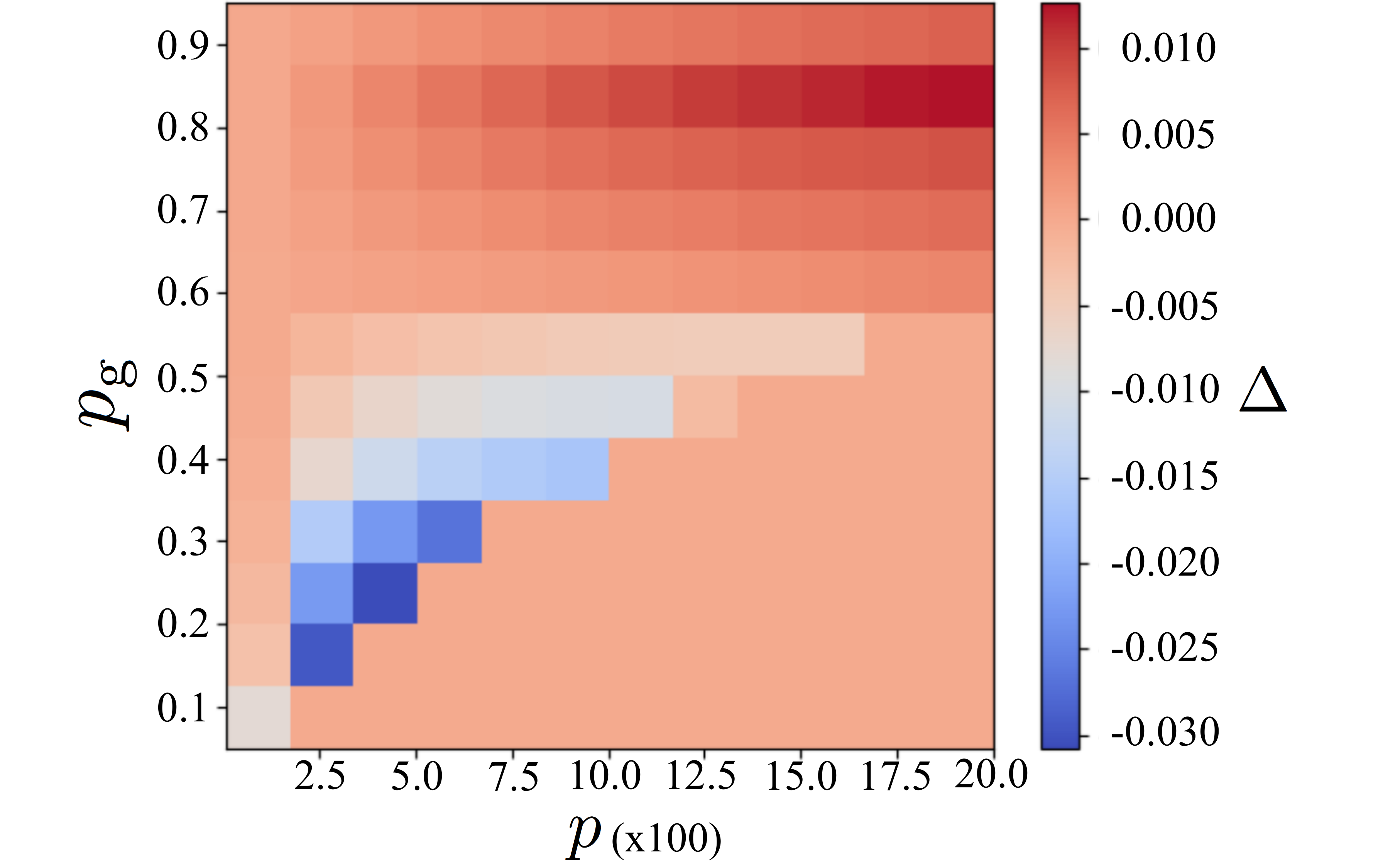}
    \caption{Difference in genuine multipartite concurrence of average state, $\Delta = C_g(\bar{\rho})_{\mathcal{E}^1} - C_g(\bar{\rho})_{\mathcal{E}^2}$, between the  $\mathcal{E}^2_{\rm M-QR}$ and the $\mathcal{E}^1_{\rm M-QR}$ under ASAP scheduling with probabilistic link generation, as a function of the per-attempt link success probability $p_{\rm g}$ and the memory depolarizing rate (denoted $p$). Blue regions ($\Delta<0$) indicate $\mathcal{E}^2_{\rm M-QR}$ performs better; red regions ($\Delta>0$) indicate $\mathcal{E}^1_{\rm M-QR}$ performs better.}
    \label{fig:triangle_prob}
\end{figure}

\paragraph*{Probabilistic link generation.}
We next let each elementary GHZ state be generated independently with success probability $p_{\rm g}$ per attempt, so qubits accumulate waiting-time-induced depolarizing noise (with parameter $p$ per round) while stored in memory. Both the primitives, inflation and deflation are performed as soon as possible (ASAP) given the available resources: $\mathcal{E}^1_{\rm M-QR}$ requires at least two elementary GHZ states to be generated before its first BSM can be performed, whereas $\mathcal{E}^2_{\rm M-QR}$ can perform a Hadamard measurement on an auxiliary qubit as soon as a single elementary GHZ state is generated, without waiting for a second. Fig.~\ref{fig:triangle_prob} shows $\Delta = C_g(\bar{\rho})_{\mathcal{E}^1} - C_g(\bar{\rho})_{\mathcal{E}^2}$ as a function of $p$ and $p_{\rm g}$. $\mathcal{E}^2_{\rm M-QR}$ remains preferable ($\Delta<0$) in the low-to-moderate $p$ regime, while $\mathcal{E}^1_{\rm M-QR}$ becomes advantageous ($\Delta>0$) once the memory depolarizing rate grows, showing that the ASAP scheduling asymmetry between the two protocols — not just the choice of measurement basis — affects which strategy is optimal in realistic scenarios. Unlike the deterministic link-generation case, here the performance metric is obtained by first averaging the output density matrices over Monte Carlo realizations and measurement outcomes, including the corresponding correction operations, and then evaluating $C_g$ of the averaged outcome state.

Taken together, these cases represent distinct instantiations of the perfromance metric $\mathcal{F}$ in Eq.~(\ref{eq:optimisation}). Although the argmax over $\{E^{1}_{\mathrm{M\text{-}QR}},E^{2}_{\mathrm{M\text{-}QR}}\}$ is well-defined in each case, the optimal protocol depends on the chosen objective and noise regime, and changes with the noise parameters. Thus, even for the smallest nontrivial topology, the optimal protocol is not fixed by topology alone, motivating the general optimization framework for larger networks and protocol spaces.

\section{Conclusion}
We developed an operational framework for the manipulation of genuine multipartite entanglement in quantum networks by classifying the relevant networking operations into three broad primitives—inflation, deflation, and purification. These SLOCC-based primitives can be composed to synthesize a broad class of first-generation M-QR protocols, providing a unified framework for protocol design, optimization, and analysis. As a representative application, we use the DCT family to demonstrate how distinct classes of multipartite entanglement can be generated and transformed in long-range distributed states in a controlled manner. We illustrated the use of the presented framework to systematically optimize multipartite quantum repeater protocols for a given network topology. Our results highlight the effectiveness of the primitive-based framework to identify optimal multipartite repeater protocols based on chosen figures of merit and repeater operating regimes in addition to the dependence on network topology.

Several directions follow naturally from this work. First, the optimization can be extended beyond the small protocol spaces considered here via systematic or heuristic search over larger topologies and richer primitive sets such as quantum error correction and multiplexing, into the primitive set. Second, the primitive-based formulation may also be extended for developing an adaptive multipartite entanglement percolation strategy, in which distinct series and parallel composition rules can be selected according to the network topology, link characteristics, and operational constraints, rather than imposing a fixed protocol throughout the network \cite{MEpercolation2010,Acin2007Percolation,concPT2021}. This also suggests a formulation of multipartite entanglement routing as a joint optimization over paths, entanglement transformations, and protocol sequences, extending existing approaches to entanglement routing and multiparty entanglement distribution \cite{routing1,routing2}. Finally, incorporating experimentally relevant costs—including memory requirements, gate and measurement operations, link-generation attempts, and finite coherence times—into the optimization objective would enable the design of resource-aware multipartite repeater protocols for near-term quantum networks.

\section{Methods}

\subsubsection*{Monte Carlo simulations for probabilistic link generation in the triangular network}

We evaluate the triangular M-QR protocols under probabilistic link generation using Monte Carlo simulations that account for stochastic generation times and the resulting memory-induced decoherence. Each elementary GHZ resource is generated independently, with
\begin{equation}
g_i\sim\mathrm{Geom}(p_i),
\end{equation}
where $g_i$ is the number of generation rounds required for resource $i$ to become available. The primary setting is the symmetric case $p_i=p_{\rm g}$, while asymmetric generation probabilities can also be considered.

For each realization, the sampled generation times determine the measurement schedule of the protocol. A qubit that remains idle for $r$ rounds is subjected to an effective depolarizing channel with
\begin{equation}
p_{\mathrm{eff}}(r)
=\frac{3}{4}\left[1-\left(1-\frac{4p}{3}\right)^r\right],
\end{equation}
where $p$ is the single-round depolarizing probability. Thus, the noisy elementary GHZ states and their subsequent fusion depend on the realization-specific waiting times and measurement order. After applying the appropriate branch corrections, we obtain the corrected output state $\rho_k$ for each realization.

The ensemble-averaged output state is obtained by averaging over both Monte Carlo realizations and measurement outcomes,
\begin{equation}
\bar{\rho}
=\frac{1}{N_{\mathrm{MC}}}\sum_{k=1}^{N_{\mathrm{MC}}}
\sum_m p_{m|k},\rho_{k,m}^{\mathrm{corr}},
\qquad
N_{\mathrm{MC}}=5000,
\end{equation}
where $p_{m|k}$ is the probability of measurement outcome $m$ in realization $k$, and $\rho_{k,m}^{\mathrm{corr}}$ is the corresponding output state after the corrective operation. The resulting $C_g(\bar{\rho})$ is used as the performance metric.

\begin{acknowledgments}
This work was supported by JST LOTUS Programme Grant Number JPMJLP25A4 as as well as by Japan’s Council for Science, Technology and Innovation (CSTI) under the Cross-ministerial Strategic Innovation Promotion Program (SIP) for “Promoting the application of advanced quantum technology platforms to social issues” (Grant JPJ012367).
\end{acknowledgments}

\bibliography{references}

\clearpage

% \onecolumngrid      % APS command to switch to single column

\clearpage
\onecolumngrid

\begin{center}
{\Large\bf Supplemental Material}\\[1em]
for\\[0.5em]
{\bf Optimising multipartite entanglement manipulation in quantum networks}
\end{center}

\section*{Examples of Inflationary operations}
\label{app:iinflation operations}
Inflation operations join neighboring multipartite states at a network node by performing measurements or unitary interactions on selected qubits from each state. This includes entanglement swapping, merging operations, and entangling unitaries. These three operations differ in the local operators \({\hat O}^{\mu_i}_{\rm I}\), the associated outcome ranges for \(\mu_i\) and the support of the post-inflation state. 

\textit{1. Entanglement swapping:}
It establishes correlations between distant parties by performing a joint Bell-state measurement on qubits belonging to neighboring multipartite units. Let $\{\ket{\Phi^{(i)}}\}_{i=0}^{3}$ denote the standard two-qubit Bell basis. A Bell-state measurement (BSM) on qubits $a\in Q_i$ and $b\in Q_{i+1}$ is implemented by the projectors
\[
\Pi_{ab}^{(r)}=\ket{\Phi^{(r)}}\bra{\Phi^{(r)}}_{ab}, \qquad r\in\{0,1,2,3\}.
\]
For BSMs applied across units $i=1,\dots,N$, the global measurement operator in Eq.~\eqref{eq:swap-squeezed} is
\[
\bigotimes_{i=1}^{N-1} {\hat O}^{\mu_i}_{\rm I}
=
\bigotimes_{i=1}^{N-1} \Pi_{a_i b_i}^{(\mu_i)} .
\]
Each $\mu_i\in\{0,1,2,3\}$, yielding $4^{\,N-1}$ possible outcome strings $\boldsymbol{\mu}$. The conditional residual state $\rho^R_{\boldsymbol{\mu}}$ follows from Eq.~\eqref{eq:swap-squeezed}.

\textit{2. Merging:}
Merging combines neighboring multipartite units by applying a controlled-NOT (CNOT) gate between selected qubits followed by a local measurement. Specifically, a merge between qubits $a\in Q_i$ and $b\in Q_{i+1}$ consists of a CNOT with control $a$ and target $b$, followed by a computational-basis measurement of $b$. The corresponding Kraus operators are
\[
M_{ab}^{(0)}=(\ket{0}\bra{0}_b)\,{\rm CNOT}_{a\to b},~
M_{ab}^{(1)}=(\ket{1}\bra{1}_b)\,{\rm CNOT}_{a\to b}.
\]
For merging operations on units $i=1,\dots,N$, the global operator is
\[
\bigotimes_{i=1}^{N-1} {\hat O}^{\mu_i}_{\rm I}
=
\bigotimes_{i=1}^{N-1} M_{a_i b_i}^{(\mu_i)},
\]
with $\mu_i\in\{0,1\}$, yielding $2^{\,N-1}$ possible outcome strings $\boldsymbol{\mu}$. The conditional post-measurement state follows Eq.~\eqref{eq:swap-squeezed}.

\textit{3. Entangling unitaries:}
Entangling unitaries connect neighboring multipartite units through unitary interactions between selected qubits. Specifically, qubits $a_i\in Q_i$ and $b_i\in Q_{i+1}$ are coupled by the unitary
\[
U_{a_i b_i}:\mathcal H_{a_i}\otimes\mathcal H_{b_i}\to\mathcal H_{a_i}\otimes\mathcal H_{b_i}.
\]
For operations applied across units $i=1,\dots,N$, the global operator factorizes as
\[
\bigotimes_{i=1}^{N-1} {\hat O}^{\mu_i}_{\rm I}
=
\bigotimes_{i=1}^{N-1} U_{a_i b_i}.
\]
Since no measurement is performed, the operation has a single deterministic outcome, and one may set $\mu_i=0$ for all units. The post-inflation state is therefore
\[
\rho'=
\Big(\bigotimes_{i=1}^{N-1} U_{a_i b_i}\Big)
\Big(\bigotimes_{i=1}^{N} \rho_{\mathcal N_i}\Big)
\Big(\bigotimes_{i=1}^{N-1} U_{a_i b_i}\Big)^\dagger ,
\]
with $P_{\boldsymbol{\mu}}=1$ for the unique outcome $\boldsymbol{\mu}=(0,\dots,0)$. Although deterministic, the inflation constraint of Eq.~\eqref{eq:general_inflation_bound} remains applicable since it follows from the monotonicity of the genuine multipartite concurrence under the SLOCC maps implementing the primitive connections.

\begin{figure}[h]
    \centering
    \includegraphics[width=0.55\linewidth]{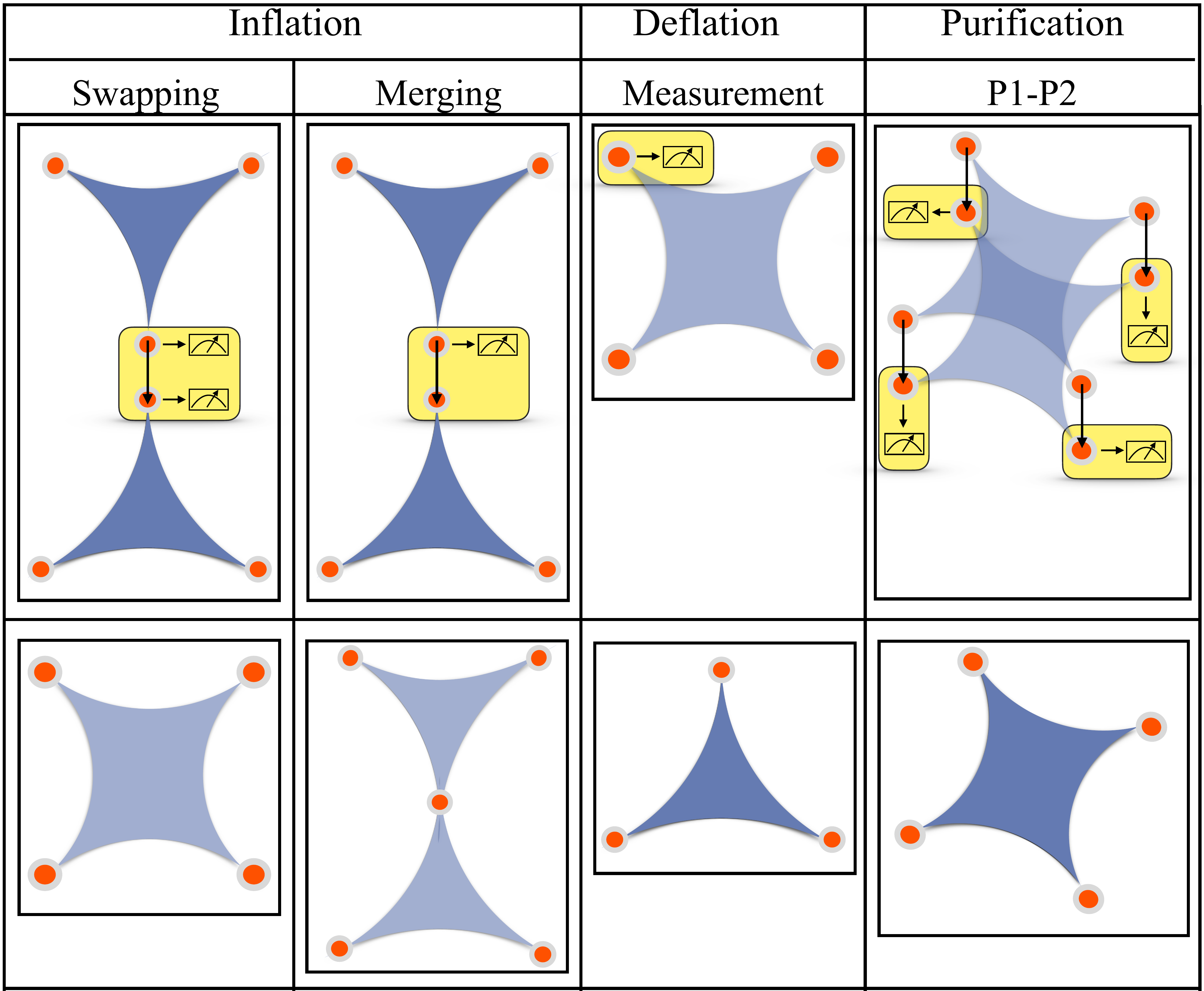}
    \caption{ Schematic representation of multipartite networking operations and their effect on the entangled states in the network. Inflation connects neighboring entangled units increasing the number of parties involved in the entangled state and reduces the GMC whereas deflation disentangle selected parties reducing the number of qubits while possibly concentrating the GMC (not necessarily in general). Purification acts on multiple copies of a multipartite state and probabilistically enhances the GMC while preserving the number of parties involved.}
    \label{fig:schematic}
\end{figure}

\FloatBarrier

\section*{GME under swapping of gGHZ and gW states}
\label{app:gGHZandGW}
\noindent
\paragraph{1. gGHZ inputs.}
Consider two generalized GHZ states
\begin{equation}
\lvert\mathrm{gGHZ}_i\rangle
= a_i \lvert 0\rangle^{\otimes N_i} + b_i \lvert 1\rangle^{\otimes N_i},
\qquad |a_i|^2 + |b_i|^2 = 1 ,
\end{equation}
for \(i=1,2\). A BSM on one qubit from each input yields, for every Bell outcome, a post-measurement state on \(N_1+N_2-2\) qubits that is again of gGHZ form, up to local Pauli corrections. As a consequence, the GMC of each conditional output factorises, and averaging over the four Bell outcomes gives the exact identity
\begin{equation}
\label{eq:appendix_gGHZ_product}
\langle C_g^{\mathrm{(out)}} \rangle
= C_g^{(1)}\, C_g^{(2)}.
\end{equation}
This strict multiplicativity follows from the full permutation symmetry of gGHZ states: all \(1{:}(N-1)\) bipartitions are equivalent and contribute identically to \(C_g\). Although the average obeys the product law, individual Bell outcomes can yield conditional states with enhanced GMC, allowing probabilistic entanglement concentration via swapping.

\noindent
\paragraph{2. gW inputs.}
For generalized W states of the form
\begin{equation}
\lvert\mathrm{gW}\rangle
= \sum_{k=1}^{N} a_k \lvert 0 \ldots 1_k \ldots 0 \rangle,
\qquad \sum_{k=1}^{N} |a_k|^2 = 1 ,
\end{equation}
the averaged post-swap GMC does not satisfy a universal product rule. In general,
\begin{equation}
\label{eq:appendix_gW_nonproduct}
\langle C_g^{\mathrm{(out)}} \rangle
\;\lessgtr\;
C_g^{(1)}\, C_g^{(2)},
\end{equation}
where the inequality direction depends on the state parameters and on which qubits are involved in the BSM. The only general constraint is the local bound for one swap is
\begin{equation}
\label{eq:appendix_local_bound}
\langle C_g^{\mathrm{(out)}} \rangle
\le C_1^{\ell}\, C_2^{\ell}.
\end{equation}

\textit{Reason of the contrast.}
The product law \eqref{eq:appendix_gGHZ_product} for gGHZ states arises from their two-dimensional support and complete bipartition symmetry, which makes the GMC insensitive to the choice of swapped qubits. By contrast, gW states lack this symmetry: different single-qubit versus rest bipartitions carry different entanglement weights, so swapping can alter the dominant bipartition entering \(C_g\), leading to non-universal scaling. Nevertheless, as in the gGHZ case, specific Bell outcomes can probabilistically enhance \(C_g\), enabling purification-by-swapping at the conditional level.

\section*{Example: the product bound ($\prod_i C^\ell_i$) fails under iterated swapping of generalized $W$ states}
\label{sec:wchain-counterexample}

We show that the product-form bound ($\prod_i C^\ell_i$), even when restricted to entanglement swapping and applied iteratively along a chain in the natural way, does not hold beyond a single junction.

Consider three normalized tripartite generalized $W$ states,
\begin{align}
\ket{A} &= 0.7\ket{100} + 0.7\ket{010} + \sqrt{0.02}\,\ket{001}, \label{eq:wA}\\
\ket{B} &= \sqrt{0.02}\,\ket{100} + \sqrt{0.96}\,\ket{010} + \sqrt{0.02}\,\ket{001}, \label{eq:wB}\\
\ket{C} &= \sqrt{0.02}\,\ket{100} + 0.7\ket{010} + 0.7\ket{001}, \label{eq:wC}
\end{align}
supported on qubits $(A_1A_2A_3)$, $(B_1B_2B_3)$, and $(C_1C_2C_3)$ respectively, with basis states listed in the order $(\ket{100},\ket{010},\ket{001})$. Direct computation of the three single-qubit reduced concurrences of each state gives
\begin{equation}
C_g(A) = C_g(B) = C_g(C) = 0.28,
\end{equation}
with the minimum in each case realized on qubits $A_3$, $B_1$ (equivalently $B_3$, by the symmetry of $\ket{B}$), and $C_1$ --- i.e., \emph{every} connector concurrence ($C^\ell$) entering the bound, including both ports of the middle resource $B$, equals exactly $0.28$. This removes any ambiguity about which port of $B$ is meant to enter the product.

We connect the three units into a chain by two entanglement-swapping (Bell-state measurement) steps, first on $(A_3,B_1)$ and then on $(B_3,C_1)$, retaining the five parties $A_1,A_2,B_2,C_2,C_3$. Direct contraction of the two Bell measurements with the nine-qubit state $\ket{A}\otimes\ket{B}\otimes\ket{C}$, followed by exact evaluation of the genuine multipartite concurrence (minimum over all $2^{4}-1=15$ bipartitions) of the retained five-qubit pure state on each of the $4\times4=16$ outcome branches, gives the exact outcome-averaged value
\begin{equation}
\sum_{\mu} P_\mu\, C_g(\psi_\mu) = 0.0411091591272\ldots \;>\; 0.28^3 = 0.021952, \label{eq:exact-chain-value}
\end{equation}
violating niave product bound ($C^\ell_1C^\ell_2C^\ell_3$) by nearly a factor of two. The branch-resolved values, grouped by Bell-outcome parity (each row combining the four sign choices $\pm,\pm$, which are related by local Pauli corrections and share a common probability and GMC), are given in Table~\ref{tab:wchain}.

\begin{table}[h]
\centering
\caption{Outcome branches for the two-step swapping chain of Eqs.~\eqref{eq:wA}--\eqref{eq:wC}, grouped by Bell-measurement parity at the two junctions. Each row represents four sign combinations ($\Phi^\pm,\Psi^\pm$) related by local corrections.}
\label{tab:wchain}
\begin{tabular}{@{}lcc@{}}
\toprule
Bell parity pair & Probability per branch & $C_g$ per branch \\
\midrule
$\Phi,\Phi$ & $0.230692$ & $0.041213025780$ \\
$\Phi,\Psi$ & $0.009508$ & $0.029003788226$ \\
$\Psi,\Phi$ & $0.009508$ & $0.029003788226$ \\
$\Psi,\Psi$ & $0.000292$ & $0.747391786828$ \\
\bottomrule
\end{tabular}
\end{table}

This example isolates a single mechanism: the connector qubit $B_3$ entering the second swap is, after the first swap, no longer in a state independent of $A_1,A_2$ --- precisely the assumption a single-junction product bound requires. Naively re-applying the single-junction product bound at each step of a chain, treating its output as a fresh independent input for the next step, is therefore not justified in general, and Eq.~\eqref{eq:exact-chain-value} shows explicitly that it is not merely non-tight but violated.

\section*{Multipartite entanglement purification}
\label{app:GME_purification}
We describe an entanglement purification primitive which directly operates on the multipartite entangled states, P1-P2 protocol, and was first introduced in \cite{Murao1998MEP5}. Several adaptive version of the P1-P2 have been developed, recognizing that the optimal sequence of applying P1 and P2 can vary depending on the initial state, allowing for more efficient enhancement of fidelity. Different variants of the entanglement purification protocol correspond to distinct strategies for choosing the sequence of purification steps, P1 and P2, has been proposed.

\begin{figure}
    \centering
    \includegraphics[width=0.5\linewidth]{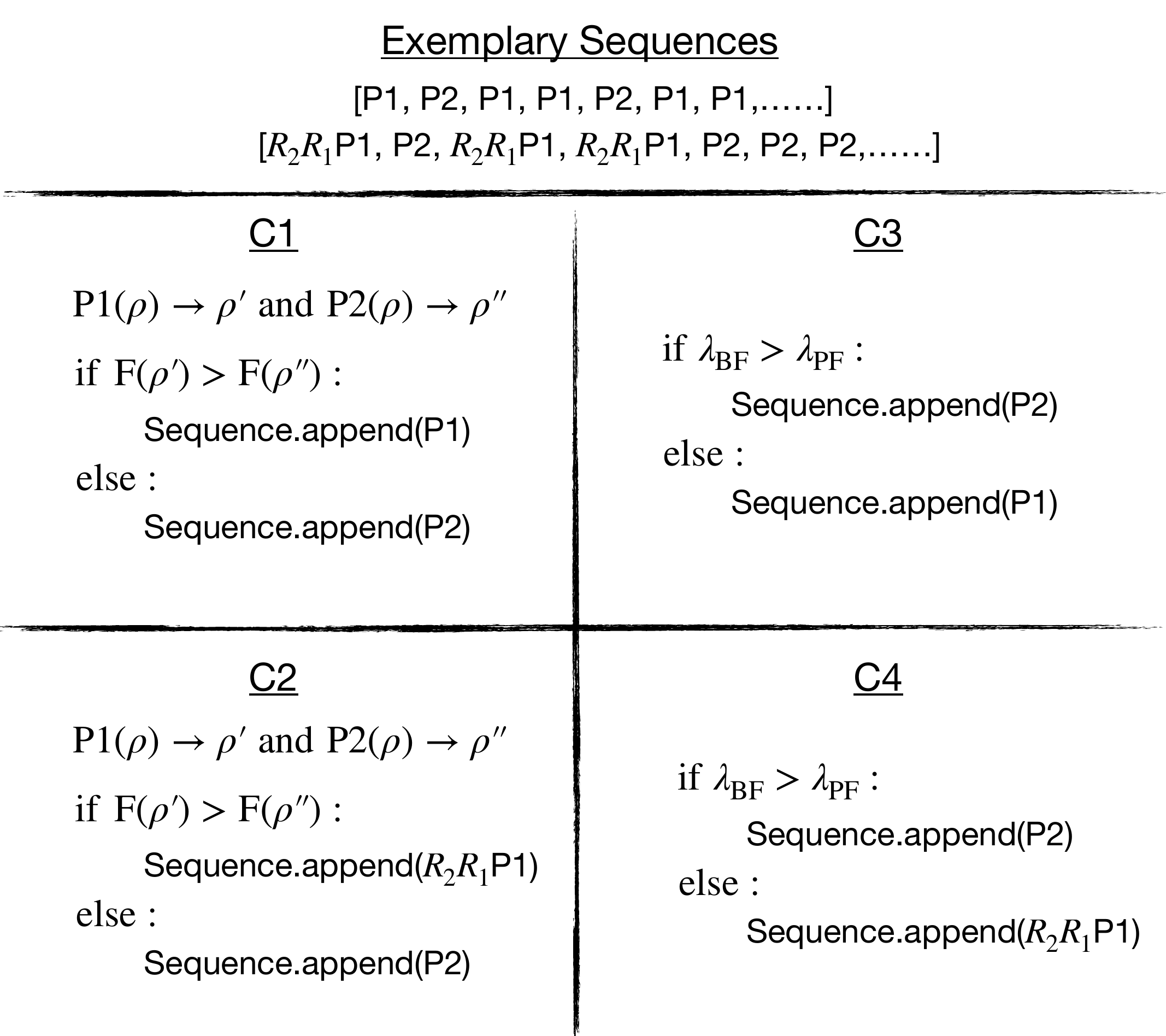}
    \caption{Pseudocode for the four MEP protocol variants (C1–C4), each of which precomputes an optimal sequence of P1 and P2 operations based on distinct selection criteria: C1 decides P1 or P2 depending which maximizes the fidelity at each step, C2 mirrors C1 applies additional depolarization following P1, C3 chooses between P1 and P2 according to bit‑flip ($\lambda_{\rm BF}$) versus phase‑flip ($\lambda_{\rm PF}$) error weights, and C4 combines weight‑based selection with depolarization. Two example purification sequences produced by these algorithms are illustrated at the top of the figure.}
    \label{fig:pseudo_codes}
\end{figure}

Here we talk about four such variants which we denote by C1, C2, C3, and C4.  The pseudocode for the four variants of P1-P2 protocols is presented in Fig.~\ref{fig:pseudo_codes}. The four adaptive multipartite entanglement purification schemes are - C1 implements a maximum local fidelity adaptive scheme, in which, at each iteration, the protocol (either P1 or P2) that leads to the higher fidelity is selected. This approach ensures a monotonic increase in fidelity with every iteration. In contrast, C3 is designed based on the observation that P2 primarily corrects bit-flip errors, while P1 addresses phase-flip errors. Accordingly, if the bit-flip error weight exceeds the phase-flip error weight, P2 is applied; otherwise, P1 is used. We introduce two additional variants, C2 and C4, which mirror C1 and C3 respectively but include one key modification: each time P1 is applied, an extra depolarizing operation is performed. These depolarizing operators preserve the diagonal elements while eliminating all anti-diagonal terms(in the computational basis) except those corresponding to GHZ . The set of depolarizing operators \cite{Epping2017NQKD}, \(\mathbb{D}\), is defined as
\begin{align}
\mathbb{D} = \{R_k \mid k = 1, 2,\dots,N-1 \},
\end{align}
with
\begin{align}
R_k &= \text{diag}(1,\, i)_1 \otimes \text{diag}(1,\, -i)_{k+1}.
\end{align}

Each party applies these operators with a probability of \(1/2\); otherwise, the identity \(\mathbb{I}\) is applied.

\section*{GME of DCT states under inflation}
\label{app:GME_inflation}

\textit{Lower bound}--Consider two DCT states with geometric measure of genuine multipartite entanglement (GMC)
\(C_g^{(1)}\) and \(C_g^{(2)}\).
Using the DCT parametrization, the post-inflation averaged GMC after one swapping step is
\begin{equation}
C_g^{\mathrm{(out)}}(\rho^R)
=
\max\{0, 2(\lambda_c^1+\lambda_0^1)(\lambda_c^2+\lambda_0^2)-1+\mu\},
\end{equation}
where \(\mu = (2\lambda_0^{(1)}\lambda_0^{(2)} + 4\lambda_*^{(1)}\lambda_*^{(2)})\ge0\) and it depends on the coefficient sets \(\{\lambda_i^1\}\) and \(\{\lambda_i^2\}\).
Since \(C_g^{(k)}=\max\{0, 2(\lambda_c^k+\lambda_0^k)-1\}\), this yields the bound
\begin{equation}
C_g^{\mathrm{(out)}}(\rho^R)
\ge
\max\left\{0,\frac{(C_g^{(1)}+1)(C_g^{(2)}+1)}{2}-1\right\}
\end{equation}
Iterating this inequality for \(N\) units gives the lower bound in Eq.~\eqref{eq:Cgmelb_rewritten}.\hfill\(\square\)

\textit{Upper bound}--To establish the upper bound on post swap GMC, it suffices to show that $\big\langle  C_g^{(1)}\,C_g^{(2)} -C_g^{\mathrm{(out)}}\big\rangle \geq 0$.  In terms of the DCT-state coefficients this condition reduces to
\begin{equation}
(1-\lambda_c^{(1)}-\lambda_0^{(1)})(1-\lambda_c^{(2)}-\lambda_0^{(2)})
-\lambda_0^{(1)}\lambda_0^{(2)}
-2\lambda_*^{(1)}\lambda_*^{(2)} \ge 0,
\label{eq:DCT_reduced_ineq}
\end{equation}
where \(\lambda_*\) denotes the coefficient of the basis state mapped, upon measurement, to the symmetric superposition
\((\ket{\Psi_{N-1}^{0+}}+\ket{\Psi_{N-1}^{0-}})/\sqrt{2}\). Normalization of each DCT state implies
\(
\lambda_c+2\lambda_0+2\lambda_* \le 1,
\)
and hence
\(
1-\lambda_c-\lambda_0 \ge \lambda_0+2\lambda_*.
\)
Applying this bound to both input states yields
\[
(1-\lambda_c^{(1)}-\lambda_0^{(1)})(1-\lambda_c^{(2)}-\lambda_0^{(2)})
\ge
(\lambda_0^{(1)}+2\lambda_*^{(1)})(\lambda_0^{(2)}+2\lambda_*^{(2)}),
\]
whose expansion is larger than
\(
\lambda_0^{(1)}\lambda_0^{(2)}+2\lambda_*^{(1)}\lambda_*^{(2)}
\)
since all coefficients are nonnegative.  
This proves Eq.~\eqref{eq:Cgmelb_rewritten} and hence the desired bound.\hfill\(\square\)

Notably for DCT states, different  measurement outcomes corresponding to swapping operation, are locally convertible by LOCC and have the same $C_g$, so reporting $C_g$ for any single outcome is equivalent to reporting the outcome-averaged value.

\section*{GME of DCT states under deflation}
\label{app:GME_deflation}

Let \(\rho\) be an \(N\)-qubit D\"ur--Cirac--Tarrach (DCT) state, which is diagonal in the GHZ basis. Consider measuring one qubit in the rotated basis
\(\{\ket{0_\theta},\ket{1_\theta}\}\), producing post-measurement states
\(\rho_\theta^\pm\) with probabilities \(p_\pm\).
\begin{equation}
\ket{0_\theta}
=\cos\theta\ket0+\sin\theta\ket1,\qquad
\ket{1_\theta}
=-\sin\theta\ket0+\cos\theta\ket1.
\end{equation}
The corresponding unnormalized post-measurement operators are
\begin{equation}
M^{+}(\theta)
=(\bra{0_\theta}\otimes I)\rho
(\ket{0_\theta}\otimes I),
\qquad
M^{-}(\theta)
=(\bra{1_\theta}\otimes I)\rho
(\ket{1_\theta}\otimes I),
\end{equation}
with probabilities
\(
p^\pm(\theta)=\Tr M^\pm(\theta),
\)
and normalized conditional states
\(
\rho_\theta^\pm
=\frac{M^\pm(\theta)}{p^\pm(\theta)}.
\)
For DCT states, the two outcomes occur with equal probability because the recudced density matrix on any qubit is an identity.
\(
p^+(\theta)=p^-(\theta)=\frac12
\)
for every \(\theta\).

Define the fidelity of each conditional state with its own GHZ-frame reference by
\begin{equation}
F^\pm(\theta)
\equiv
\bra{\Psi_{n-1}^{0\pm}}
\rho_\theta^\pm
\ket{\Psi_{n-1}^{0\pm}},
\end{equation}
where \(\ket{\Psi_{n-1}^{0-}}\) is related to
\(\ket{\Psi_{n-1}^{0+}}\) by a local \(Z\) operation. The two fidelities are equal and are given by
\begin{equation}
F^+(\theta)=F^-(\theta)\equiv F(\theta)
=\frac{\lambda_c}{2}
+\lambda_0+\lambda_\ast
+\frac{\lambda_c}{2}\sin 2\theta .
\label{eq:Ftheta}
\end{equation}

Since \(p^+=p^-=\frac12\) and \(F^+(\theta)=F^-(\theta)\), the outcome-averaged GME concurrence is
\begin{align}
A(\theta) = 
p^+C_g(\rho_\theta^+)
+p^-C_g(\rho_\theta^-)
=
\max\{0,\lambda_c t -2 \sum_j \sqrt{4b_jd_j + t^2(b_j-d_j)^2}\} \equiv A(t)
,
\label{eq:Adef}
\end{align}
where, $t=sin(2\theta)$ and $\{b_j,d_j\}$ denotes the pairs of complementary non-corner diagonal weights of the pre-measurement $(n-1)$-qubit conditional block obtained by fixing the measured qubit.

Now for any $b,d>0$ and $t \in [0,1]$,
\[
4bd+ t^2(b-d)^2 \geq t^2(b+d)^2
 \]

Therefore, the un-clipped value of average GMC satisfies, 
\[A(t) \leq tA(1) \leq A(1) \qquad \forall t \in [0,1] \]
with equality iff $t=1$. This establishes $
\theta =\pi/4$ uniquely maximizes the outcome averaged GMC over single-qubit projective measurement.

% where
% \(
% D\equiv 2(\lambda_c+\lambda_0)-1,
% \quad
% C_g(\rho)=\max\{0,D\}.
% \)

% To optimize \(A(\theta)\), define
% \begin{equation}
% g(\theta)
% =D+2\lambda_\ast+\lambda_c(\sin2\theta -1) .
% \end{equation}
% For the DCT coefficients, \(\lambda_c\ge0\), and hence
% \begin{equation}
% g'(\theta)=2\lambda_c\cos2\theta,
% \end{equation}
% whose stationary points are
% \(\theta=\pi/4\) and \(3\pi/4\) modulo \(\pi\). Moreover,
% \begin{equation}
% g''(\pi/4)=-4\lambda_c\le0,
% \qquad
% g''(3\pi/4)=4\lambda_c\ge0.
% \end{equation}
% Thus \(\theta=\pi/4\) is the global maximum and
% \(\theta=3\pi/4\) the global minimum. Equivalently,
% \(\sin2\theta\le1\) for all \(\theta\), and the monotonicity of
% \(x\mapsto\max\{0,x\}\) preserves this ordering. Therefore,

Now, 
\begin{equation}
A(1)
=\max\{0,D+2\lambda_\ast\}.
\end{equation}

Finally, to compare the optimized post-measurement value with the initial GME concurrence, we evaluate
\(A(1)-C_g(\rho)\),
\begin{equation}
A(\pi/4)-C_g(\rho)
=\max\{0,D+2\lambda_\ast\}
-\max\{0,D\}
\ge0,
\label{eq:mainineq}
\end{equation}
since \(\lambda_\ast\ge0\), showing that the Hadamard-basis measurement never decreases the outcome-averaged GME concurrence of the remaining \((N-1)\)-qubit state.
This completes the proof.
\hfill\(\square\)

\section*{Party-conserving mode M-QR protocol and entanglement-class identification for simulation}

We consider multipartite quantum repeater dynamics in the \emph{party-conserving mode} (PCM), where the number of qubits in the distributed state remains unchanged after each repeater iteration. The elementary resource states are taken to be of the DCT form $\rho(\lambda_c,{\lambda_i})$ introduced in Eq.~(\ref{eq:elementaryunits}). A single PCM iteration consists of entanglement swapping via a Bell-state measurement (BSM) performed on one qubit from each of two neighboring elementary units, followed by local single-qubit measurements in the Hadamard basis. Iterating this procedure generates a sequence of DCT states whose entanglement properties can be tracked analytically.

To classify the resulting states, we partition the DCT parameter space into the four standard three-qubit entanglement classes: fully separable, biseparable, W-type, and GHZ-type. GHZ-type states are identified by a nonvanishing three-tangle,
\[
\tau_3>0,
\]
whereas W-type states satisfy
\[
\tau_3=0,\qquad C_g>0.
\]
Within the DCT family, full separability is equivalent to positivity under all partial transpositions (PPT). Biseparable states are distinguished from fully separable states by the conditions
\[
\tau_3=0,\qquad C_g=0,
\]
while remaining non-PPT with respect to at least one bipartition~\cite{1999DCTintro,Eltschka2012,Siewert2012}. The classification procedure employed in the numerical simulations therefore consists of: (i) evaluating the PPT conditions for all bipartitions, (ii) computing the genuine multipartite connectivity $(C_g)$, and (iii) evaluating the three-tangle $(\tau_3)$. For the present family, the available closed-form result for $\tau_3$ is only a lower bound \cite{Eltschka2012,Siewert2012}. Consequently, it cannot be used to uniquely identify the GHZ and W regions. The region $S_2$ in the left panel of Fig.~\ref{fig:Cgme evolution tripartite 4} therefore comprises states that may belong to either the GHZ or W class.

The resulting flow through the DCT parameter space under successive PCM iterations is shown in the left panel of Fig.~\ref{fig:Cgme evolution tripartite 4}. The effect of incorporating end-to-end purification after each PCM round is shown in the right panel, where repeated purification enlarges the GHZ-entangled region $(S_1)$.

\section*{Motivation for optimisation of M-QR protocols}

\subsection*{Two M-QR implementations on the plaquette network.}

\begin{figure}[h]
    \centering
    \includegraphics[width=0.5\linewidth]{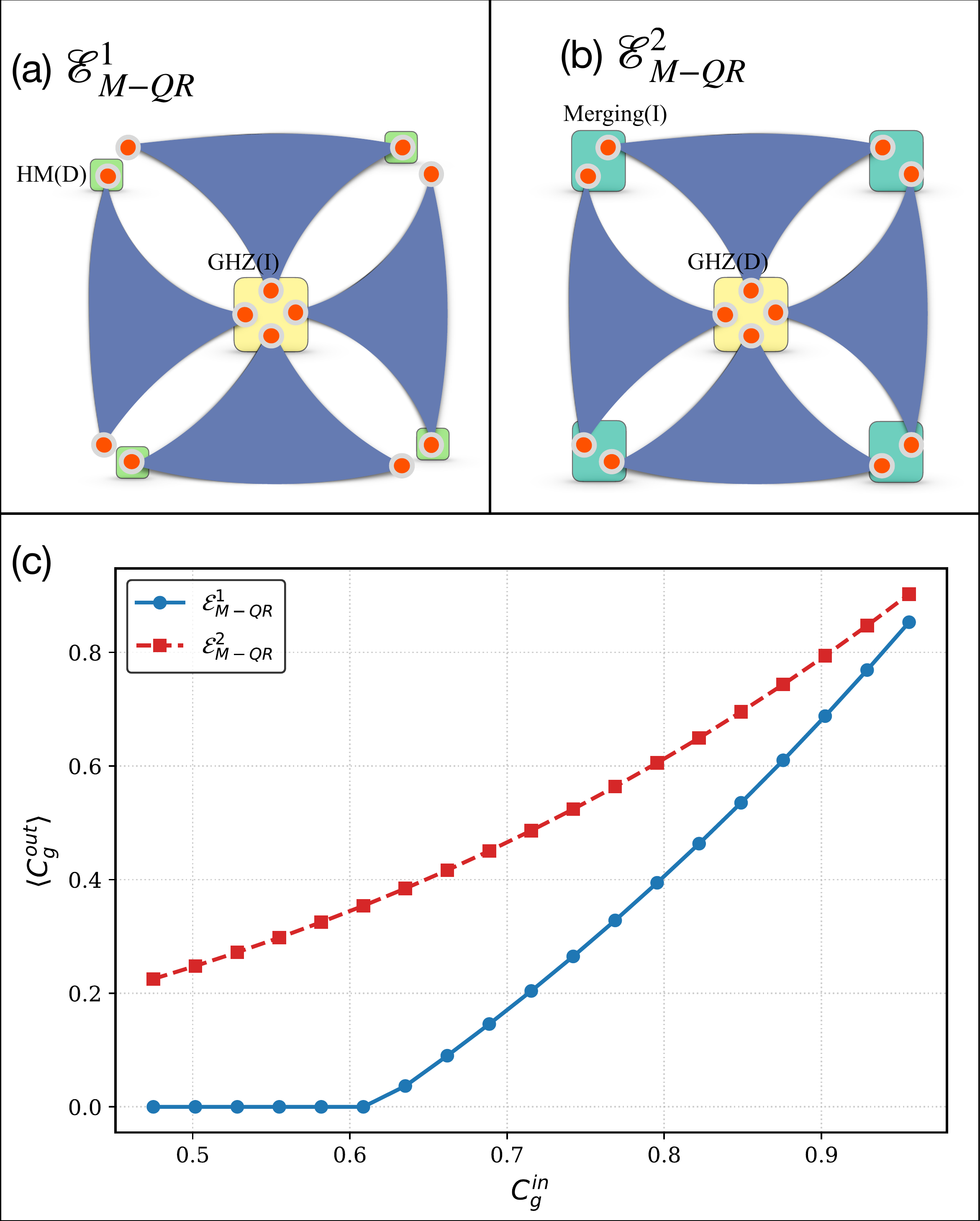}
    \caption{
Comparison of two multipartite quantum routing protocols for generating a four-party target state on the corner nodes from four triangulated elementary resource states. (a) Protocol $\mathcal{E}_{\mathrm{M\text{-}QR}}^{1}$: a central four-qubit GHZ measurement acts as the inflation operation, followed by four local Hadamard measurements. (b) Protocol $\mathcal{E}_{\mathrm{M\text{-}QR}}^{2}$: four merging operations first inflate the network into a larger multipartite state, followed by a central GHZ measurement acting as the deflation operation. (c) Average output genuine multipartite concurrence $\langle C_g^{\mathrm{out}} \rangle$ as a function of the elementary-state concurrence $C_g^{\mathrm{in}}$. Despite the emergence of unsuccessful branches in $\mathcal{E}_{\mathrm{M\text{-}QR}}^{2}$, its average output concurrence exceeds that of $\mathcal{E}_{\mathrm{M\text{-}QR}}^{1}$ throughout the investigated parameter range.
}
    \label{fig:postselection-a}
\end{figure}

We now consider the four triangulated elementary states ($\{A, B, C, D\}$) arranged in the plaquette geometry of Fig.~\ref{fig:postselection-a}, with the target four-party output supported on the corner qubits (say $\{A_1,B_1,C_1,D_1\}$). In this setting, we analyse two distinct multipartite quantum routing protocols are available, denoted by $\mathcal{E}_{\mathrm{M\text{-}QR}}^{1}$ and $\mathcal{E}_{\mathrm{M\text{-}QR}}^{2}$. For simulations we consider the elementary units to be of DCT form with just one parameter $\lambda_c$ and all the other $\lambda_i$ are same and are fixed by normalization condition.

The first protocol, $\mathcal{E}_{\mathrm{M\text{-}QR}}^{1}$, begins with a four-qubit GHZ measurement, which acts as the inflation step by fusing the relevant branches into a larger intermediate state. This is followed by four Hadamard measurements, which play the role of deflation and remove the redundant qubits, leaving the final state on $\{A_1,B_1,C_1,D_1\}$. Owing to the GHZ-diagonal structure of the intermediate states, all measurement outcomes in this protocol are related by local corrections; hence, every branch belongs to the same LOCC class and yields the same value of the output GMC up to local unitaries.

In contrast, the second protocol, $\mathcal{E}_{\mathrm{M\text{-}QR}}^{2}$, starts with four merging operations as the inflation stage and then applies a central GHZ measurement as the deflation stage. Here the inflation step initially creates a richer intermediate structure, but the subsequent deflation breaks the LOCC-equivalence among the measurement branches. As a result, the outcomes split into two qualitatively different classes: one class leads to nonzero output GMC, while the complementary class produces vanishing GMC after deflation. 

\begin{figure}[h]
    \centering
    \includegraphics[width=0.5\linewidth]{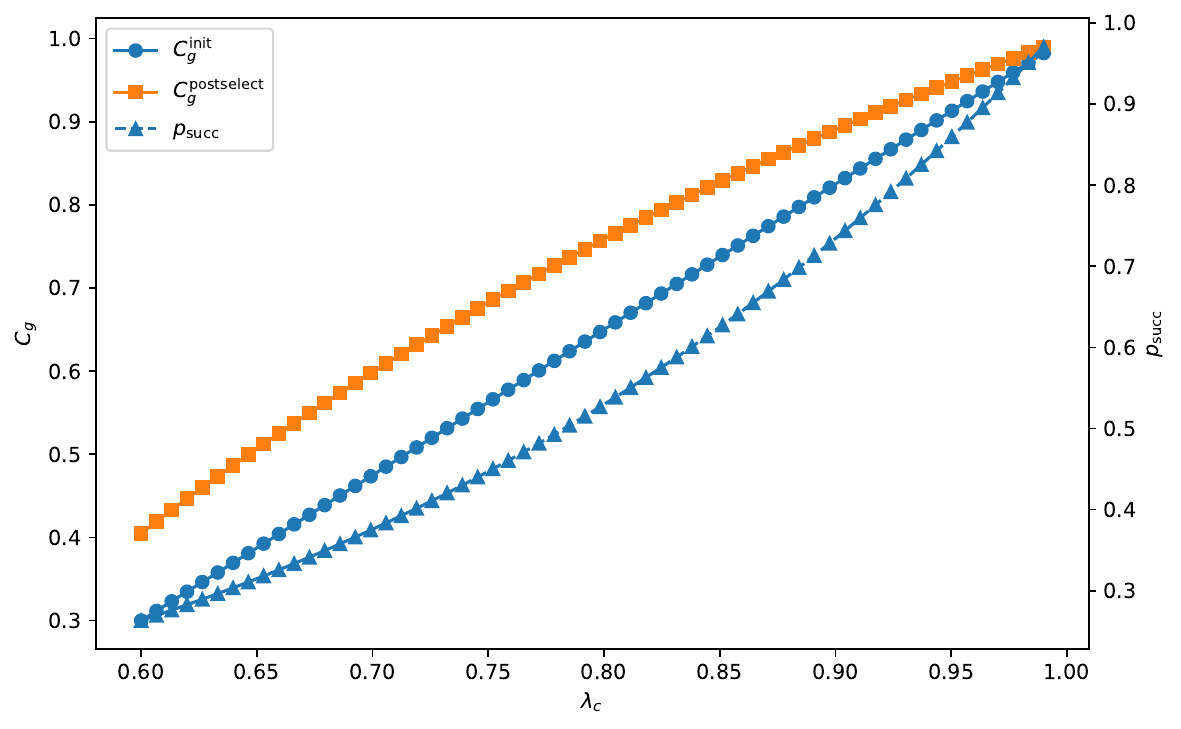}
    \caption{
Postselection-assisted multipartite quantum routing using $\mathcal{E}_{\mathrm{M\text{-}QR}}^{2}$. The measurement outcomes of the final deflation step separate into distinct classes, allowing postselection on the favorable branches that yield nonzero output GMC. The resulting postselected output concurrence $C_g^{\mathrm{out}}$ is enhanced relative to the averaged protocol performance and exhibits a reduced effective error threshold. Remarkably, for a broad range of elementary-state parameters, the postselected branch produces an output concurrence exceeding the concurrence of the elementary resource states themselves. This reflects conditional concentration of GMC in the postselected branch, combined with the spatial extension of the entangled state. The corresponding success probability remains sufficiently high over a significant region of parameter space, making the postselection strategy operationally relevant.
}
    \label{fig:postselection}
\end{figure}

\begin{itemize}
    \item Class A (outcomes in $S_+$, cumulative probability $P_+$): post-deflation states with $C_g(\rho^{\mathrm{out}}) > 0$.
    \item Class B (outcomes in $S_-$, probability $P_- = 1 - P_+$): post-deflation states with $C_g(\rho^{\mathrm{out}}) = 0$.
\end{itemize}
Importantly, the Class~A outcomes are themselves \emph{LOCC-interconvertible within their class}: the post-selected output is a unique (up to local unitaries).The Fig.~(\ref{fig:postselection}) shows the result of the simulation.

\FloatBarrier

This loss of uniform LOCC interconvertibility is precisely what distinguishes $\mathcal{E}_{\mathrm{M\text{-}QR}}^{2}$ from $\mathcal{E}_{\mathrm{M\text{-}QR}}^{1}$. Nevertheless, as shown in Fig.~\ref{fig:postselection-a}, the average output $\langle C_g^{\mathrm{out}}\rangle$ generated by $\mathcal{E}_{\mathrm{M\text{-}QR}}^{2}$ remains larger than that of $\mathcal{E}_{\mathrm{M\text{-}QR}}^{1}$ over the plotted range of initial $C_g$, even though part of the probability mass is concentrated in zero-GMC branches.

Moreover, $\mathcal{E}_{\mathrm{M\text{-}QR}}^{2}$ also admits a useful postselection strategy on favorable measurement outcomes. Conditioning on these outcomes further enhances the output GMC and lowers the effective error threshold, as indicated by the corresponding performance curve in Fig~\ref{fig:postselection}. In particular, the postselected branch can amplify the GMC of the inflated state beyond the initial $C_g$ of the elementary units, while still retaining a sufficiently high success probability.

\FloatBarrier

\end{document}